\documentclass{article}
\usepackage{arxiv}
\usepackage{cite}
\usepackage{amsmath,amssymb,amsfonts}
\usepackage{graphicx}
\usepackage{algorithm,algorithmic}
\usepackage{hyperref}
\usepackage{textcomp}
\usepackage{color,soul}
\usepackage{microtype}
\usepackage{nicefrac}     
\usepackage{subcaption}
\usepackage{makecell}
\usepackage{pifont}
\usepackage{placeins}
\usepackage{tabularx}
\usepackage{stmaryrd}

\usepackage[utf8]{inputenc} 
\usepackage[T1]{fontenc}    
\usepackage{url}            
\usepackage{booktabs}       
\usepackage{nicefrac}       
\usepackage{microtype}      
\usepackage{doi}
\SetSymbolFont{stmry}{bold}{U}{stmry}{m}{n}
\def\BibTeX{{\rm B\kern-.05em{\sc i\kern-.025em b}\kern-.08em
    T\kern-.1667em\lower.7ex\hbox{E}\kern-.125emX}}

\newcommand{\R}{\mathbb{R}}
\newcommand{\C}{\mathcal{C}}
\newcommand{\Set}{{\mathcal{S}(T;u^*)}}
\newcommand{\N}{\mathcal{N}}
\definecolor{codegreen}{rgb}{0,0.6,0}
\newcommand{\q}{\boldsymbol{q}}
\newcommand{\qd}{\dot{\boldsymbol{q}}}
\newcommand{\qdd}{\ddot{\boldsymbol{q}}}
\newcommand{\dq}{\boldsymbol{q}_r}
\newcommand{\dqd}{\dot{\boldsymbol{q}}_r}
\newcommand{\dqdd}{\ddot{\boldsymbol{q}}_r}

\newcommand{\err}{\boldsymbol{e}}
\newcommand{\ed}{\dot{\boldsymbol{e}}}
\newcommand{\eps}{\boldsymbol{\varepsilon}}
\newcommand{\epsd}{\dot{\boldsymbol{\varepsilon}}}

\newcommand{\norm}[1]{\lVert #1 \rVert}
\newcommand{\abs}[1]{| #1 |}
\newcommand{\bs}[1]{\boldsymbol{ #1 }}
\newcommand{\ol}[1]{\overline{ #1 }}
\newcommand{\udl}[1]{\underline{ #1 }}
\newcommand{\paranthesis}[1]{\left( #1 \right)}

\newcommand{\mSig}[1]{\lceil \boldsymbol{#1} \rfloor^0}
\newcommand{\Min}{\text{min}}
\newcommand{\Max}{\text{max}}

\newcommand{\cmark}{\ding{51}}
\newcommand{\xmark}{\ding{55}}

\newtheorem{theorem}{Theorem}
\newcommand{\Theorem}[2]{\begin{theorem}#2 #1\end{theorem}}

\newtheorem{Lemma}{Lemma}

\newtheorem{remark}{Remark}
\newcommand{\Remark}[2]{\begin{remark}#2 #1\end{remark}}

\newtheorem{definition}{Definition}

\newtheorem{assumption}{Assumption}
\newcommand{\assume}[2]{\begin{assumption} #2 #1\end{assumption}}

\newtheorem{property}{Property}
\newcommand{\prop}[2]{\begin{property} #2 #1\end{property}}

\newtheorem{problem}{Problem}

\newcommand{\prob}[2]{\begin{problem} #2 #1\end{problem}}

\newcommand{\assumeref}[1]{\textit{Assumption} #1}
\newcommand{\propref}[1]{\textit{Property} #1}

\newcommand{\policy}{\eqref{eq:ptcuel_ControlPolicy}}

\title{Tracking Control of Euler-Lagrangian Systems with Prescribed State, Input, and Temporal Constraints}
\author{\href{https://orcid.org/0000-0002-4270-9443}{\includegraphics[scale=0.06]{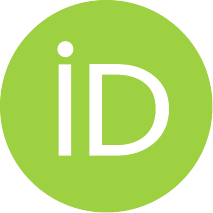}\hspace{1mm}Chidre Shravista Kashyap}  \\ {Center for Cyber-Physical Systems} \\ {IISc, Bengaluru, India} \\ {chidres@iisc.ac.in} \And \href{https://orcid.org/0000-0002-5452-8850}{\includegraphics[scale=0.06]{orcid.pdf}\hspace{1mm}Pushpak Jagtap} \\{Center for Cyber-Physical Systems} \\ {IISc, Bengaluru, India} \\ {pushpak@iisc.ac.in}  \And \href{https://orcid.org/0000-0001-8770-2301}{\includegraphics[scale=0.06]{orcid.pdf}\hspace{1mm}Jishnu Keshavan} \\{Dept. of Mechanical Engineering} \\ {IISc, Bengaluru, India} \\ {kjishnu@iisc.ac.in}}

\hypersetup{
pdftitle={Tracking Control of Euler Lagrangian Systems with SIT constraints},
pdfsubject={Control systems},
pdfauthor={Chidre Shravista Kashyap, Pushpak Jagtap, Jishnu Keshavan },
pdfkeywords={Prescribed Time Convergence, Euler Lagrangian Systems, adaptive barrier function control, state and input constraints, approximation-free control.},
}

\begin{document}

\maketitle

\begin{abstract}
The synthesis of a smooth tracking control for Euler-Lagrangian (EL) systems under stringent state, input, and temporal (SIT) constraints is challenging. In contrast to existing methods that utilize prior knowledge of EL model parameters and uncertainty bounds, this study proposes an \emph{approximation-free} adaptive barrier function-based control policy to ensure \emph{local prescribed time convergence} of tracking error under state and input constraints. The proposed approach uses smooth time-based generator functions embedded in the filtered tracking error, which is combined with a saturation function that limits control action and confines states within the prescribed limits by enforcing the time-varying bounds on the filtered tracking error. Importantly, corresponding feasibility conditions are derived pertaining to the minimum control authority, the maximum disturbance rejection capability of the control policy, and the viable set of initial conditions, illuminating the narrow operating domain of EL systems arising from the interplay of SIT constraints. Finally, the efficacy of the proposed approach is demonstrated using experimental and comparison studies.
\end{abstract}


\section{Introduction} \label{sec:introduction}
Euler-Lagrangian (EL) description of dynamics requires precise knowledge of the physical properties of rigid links such as mass, inertia, and centre of mass. Though these data are typically accessible through CAD software, they tend to be erroneous, as physical objects possess uncertainty in the body's geometry and material homogeneity in real-world scenarios. Furthermore, adhering to time domain restrictions is crucial in various robot applications, where the task objective requires to be accomplished within a stipulated time. Thus, designing controllers to achieve tracking in a user-defined settling time is of great interest as it significantly broadens its applicability to robotics. One of the popular techniques for regulating the settling time is through PID control \cite{ogata2010modern}. However, control design becomes challenging in the presence of state/input constraints and model uncertainties, which need to be accounted for during successful practical implementation. 

The notion of finite-time stability \cite{Shen_FT_OBSER} and fixed-time stability \cite{basin_FT} of autonomous systems have been introduced to tackle the problem of synthesizing a control law in the presence of temporal constraints, where the former notion guarantees convergence of solutions to the equilibrium point in some finite time for a given initial condition, and the latter concept provides convergence in a fixed time independent of the initial conditions. The stability analysis typically involves choosing a Lyapunov function candidate \cite{polykao_FT_EXPAP,polykov_FT_NON} to show uniform non-asymptotic convergence and derive a conservative estimate of the finite (or fixed) convergence time. However, the notion of the prescribed-time Stability (PTS)\cite{PTS_IMA,PTC_note_ifac,song:2023}  of the system remains unmet, which is the case of the system trajectory achieving convergence in a user-defined settling time. 

\begin{table*}
    \centering
    \caption{Qualitative comparison of the proposed study with related studies.}
    \begin{tabular}{cccccccc}
    \hline
    \hline
    {Method} & PTC & Singularity-free & Approximation-free & \makecell{Continuous\\ Control law} & \makecell{State
    \\Constraints} & \makecell{Input\\ Constraints} & \makecell{Feasibility\\ Analysis}  \\\hline
\cite{CBF_EL_IP} & \xmark & -- & \xmark & \cmark & \cmark & \cmark & \cmark \\
\cite{yang_adative,van_fault_tolerant} & \xmark & -- & \cmark & \cmark & \xmark & \xmark & \xmark\\
\cite{wu_adaptive_rbfnn, wu_fxd_neural, sun_fxd_ppf} & \xmark & -- & \cmark & \xmark & \xmark & \cmark & \xmark\\
\cite{bertingo_7dof, obuz:2025} & \cmark & \xmark & \xmark & \cmark & \xmark & \xmark & \xmark\\
\cite{PTC_TBG} & \cmark & \xmark & \cmark & \xmark & \xmark & \xmark & \xmark\\
\cite{song:2017,holloway:linear_observers:2019}& \cmark & \xmark & \xmark & \cmark & \xmark & \xmark & \xmark \\
\cite{hua:aptc:2022, hua:time_dealy:2023,krishnaMurthy:dynamicHighGain:2020}&\cmark & \xmark & \xmark & \cmark & \xmark & \xmark & \xmark \\
\cite{PTC_perturbed_EL} & \cmark & \xmark & \xmark & \cmark & \xmark & \xmark & \xmark\\
\cite{kgarg_clf} & \cmark & \cmark & \xmark & \cmark & \xmark & \cmark & \cmark\\
\cite{Huang2021-ep} & \cmark & \cmark & \xmark & \xmark & \xmark & \xmark & \xmark \\
\cite{denis:discretization:2024, yorlov:timeSpace:2022, gutierrez_exact_diff:2023,yorlov:2022:robust_differentiators, ramon:robust_obser_2022,silva:robust_sensor_noise:2024, lopez_diff:2023}& \cmark & \cmark & \xmark & -- & \xmark & \xmark & \xmark \\ 
\cite{zhao_PPC} & \cmark & \cmark & \cmark & \cmark & \cmark & \xmark & \xmark\\
\cite{sun_Neural, robustPPC_EL_PTC, PTC_EL_2020_Jian} & \cmark & \cmark & \cmark & \cmark & \xmark & \xmark & \xmark\\
\cite{PPC_BLF_EL} & \cmark & \cmark & \cmark & \cmark & partial & \cmark & \xmark\\
\cite{Huang2023} & \cmark & \cmark & \cmark & \xmark & \xmark & \cmark & \xmark\\
\cite{yeCao} & \cmark & \cmark & \cmark & \cmark & \cmark & \xmark & \xmark\\
\cite{pptc_song} & \cmark & \cmark & \cmark & \xmark & \cmark & \cmark & \xmark \\
Proposed & \cmark & \cmark & \cmark & \cmark & \cmark & \cmark & \cmark \\\hline\hline
    \end{tabular}
    \label{tab:ptcuel_qualitative_comparison}
\end{table*}

\subsection{Background}
A customary method for achieving PTS is to synthesize an appropriate time-varying gain that goes to infinity as the system tracking error converges to the origin \cite{song:2017,song:2023}. {However, prescribed-time controllers with time-varying gains can become unbounded, leading to closed-loop singularities, which cause implementation issues, including instability from discretization and sensitivity to measurement noise \cite{lopez_noise:2023}.} Alternatively, the study \cite{kgarg_qp} solves a quadratic program incorporating the user-defined settling time through a prescribed-time Control Lyapunov Function (CLF). Alternatively, state \cite{song:2017, song:2023} and time scaling techniques \cite{kgarg_clf} are employed to attain PTS, where the former study transforms the state using monotonically increasing functions that increase as time approaches the user-defined settling time and, then, the control goal is to satisfy the boundedness of the morphed signal such that it implicitly makes the original error converge to zero. The latter approach is a time transformation technique \cite{ehsan_TimeTF,kgarg_clf, bertingo_7dof, yorlov:timeSpace:2022} where the time variable goes to a preset settling time as the transformed variable goes to infinity; however, it is unclear on the evolution of system trajectories once prescribed-time convergence (PTC) has been achieved in the presence of external disturbances. An alternative approach for mitigating this issue involves the use of {smooth} time-based generator (TBG) functions \cite{TBG_Higher_order, haifeg_PTC_ADA, PTC_TBG, Huang2021-ep}, {which also avoids the singularity of control gain}, but the synthesis of a control policy using this method requires prior knowledge of the system dynamics \cite{zhang_backstepping}.

The properties of EL systems \cite{bruno_book, EL_props} are typically leveraged to deal with unknown dynamical models. {In particular, the nominal model parameters are used \cite{van_fault_tolerant, sun_USDE} while treating the model errors as a lumped uncertainty term with known bound, whereas the studies \cite{yang_adative} and \cite{xiao_UD} estimate parameters through adaptive law technique, and then the controller synthesis uses these estimates to achieve tracking in finite time.} In \cite{wu_adaptive_rbfnn, wu_fxd_neural, sun_fxd_ppf}, radial basis functions are used to approximate the model dynamics, which is coupled with a first-order sliding mode controller for guaranteeing the fixed time stability while imposing input constraints. Nevertheless, the studies \cite{sun_USDE,yu_FxdTC,xiao_UD, yang_adative, wu_adaptive_rbfnn, wu_fxd_neural, sun_fxd_ppf} fail to allow the user to provide a prescribed settling time against state and input constraints to guarantee the PTS for an unknown dynamical system. 

To mitigate this limitation, the study \cite{sun_Neural} relies on a discontinuous control policy that estimates the unknown model parameters to achieve PTC. {Alternatively, the study \cite{Huang2023} uses a time-delayed estimation of the lumped uncertainty alongside prescribed-time functions in the controller synthesis to achieve prescribed-time prescribed-bound (PTPB) convergence while accounting for the input constraints}. Similarly, studies \cite{ sun_PPC, zhao_PPC, yeCao, PPC_BLF_EL, chen:2023:eventDriven} use radial basis functions to approximate the EL dynamics and leverage prescribed performance functions to synthesize a control law that achieves PTPB convergence subject to state constraints. Despite incorporating state or input constraints coupled with temporal constraints, these studies \cite{sun_Neural, sun_PPC, Huang2023, zhao_PPC, yeCao, PPC_BLF_EL} do not consider the EL system with complete SIT constraints, which is of more practical importance.

\subsection{Contributions}

{The studies \cite{hefu:timeVarying:2023, amir:2022:timevarying,wuquan:vanishing:2024} deal with a class of non-autonomous systems for PTS. However, the proposed study focuses on a class of EL systems like robotic manipulators and quadcopters \cite{quad_EL}}. The effectiveness of this study is evaluated through a qualitative comparison with the selected studies Table \ref{tab:ptcuel_qualitative_comparison}. In this study, we propose an adaptive control strategy that guarantees PTC for EL systems, in contrast to the studies \cite{yang_adative} and \cite{sun_fxd_ppf} that ensure only finite time convergence. {The proposed control law consists of non-singular time-varying gains in contrast with the studies \cite{song:2017, krishnaMurthy:dynamicHighGain:2020, PTC_perturbed_EL} whose gains become unbounded at the user-defined settling time. On the other hand, the studies \cite{kgarg_clf, denis:discretization:2024, yorlov:timeSpace:2022,ramon:robust_obser_2022, yorlov:2022:robust_differentiators, silva:robust_sensor_noise:2024, gutierrez_exact_diff:2023, Huang2021-ep,lopez_diff:2023} show that the time-varying gains do not grow infinite as time approaches the prescribed-time. Nevertheless, these studies \cite{kgarg_clf, denis:discretization:2024, yorlov:timeSpace:2022,ramon:robust_obser_2022, yorlov:2022:robust_differentiators, silva:robust_sensor_noise:2024, gutierrez_exact_diff:2023, Huang2021-ep,lopez_diff:2023, PTC_perturbed_EL, bertingo_7dof, PTC_TBG, zhao_PPC} depend on nominal model parameters, whereas the proposed control policy is based on an adaptive barrier function that is approximation-free}. 

In particular, a filtered tracking error term is synthesized based on TBG functions that specify a user-defined settling time.  A time-varying bound on the filtered tracking error is then invoked to design the control policy that ensures the satisfaction of state constraints, unlike the studies  \cite{sun_Neural}. Additionally, the studies \cite{Huang2023, robustPPC_EL_PTC, PTC_EL_2020_Jian} that tackle PTPB convergence use performance functions to constrain position coordinates only. However, guaranteeing the performance of the full-state while adhering to full state and input constraints is not developed rigorously in earlier studies. The proposed study tackles this problem by employing a nonlinear transformation of the control input using a saturation function, thus naturally bounding associated control action compared to the study \cite{yeCao}. Although the study \cite{PPC_BLF_EL} accounts for input constraints for ensuring PTS, it implements the prescribed performance functions for partial state constraints (position coordinates only), whereas the proposed adaptive barrier control policy adheres to both full state and input constraints. {The study \cite{pptc_song} proposes the development of an anti-saturation discontinuous control policy for PTPB convergence assuming symmetric state and input constraints. In contrast, this work proposes a flexible continuous policy accommodating asymmetric constraints with different lower and upper bounds on the state/input constraints.}

{Importantly, in contrast with the studies \cite{polykao_FT_EXPAP, polykov_FT_NON, PTS_IMA, PTC_note_ifac, song:2023,song:2017, yeCao, Huang2023} on PTS under input constraints, the studies \cite{kgarg_clf,kgarg_FxTsInput:2022} demonstrate that the larger the choice of the prescribed settling time, the larger the domain of attraction for the system under given input bounds. Likewise, to avoid state constraint violations under input saturation, the controller should have sufficient authority with all possible efforts to actively restrict the state trajectories within prescribed state bounds. \cite{CBF_EL_IP}. Consequently, systems under state and input constraints require the synthesis of a viable set of initial conditions thus implying local PTS \cite{hua:aptc:2022,hua:time_dealy:2023}}. In essence, the related studies mentioned in Table \ref{tab:ptcuel_qualitative_comparison} do not (i) guarantee PTPB convergence in the presence of both state and input constraints and (ii) derive the feasibility conditions establishing the minimal control authority for achieving the tracking objective in the prescribed-time and the associated maximum disturbance handling capability of the proposed scheme along with a viable/feasible set of initial conditions that can be driven to the prescribed-bound set. Altogether, the proposed study addresses the challenges of synthesizing controllers {with bounded time-varying gains} for uncertain systems subjected to SIT constraints and also provides the feasibility conditions to tackle the problem of conservative prescribed-time control \cite{song:2023}.

To the best of the author's knowledge, this is the first study that considers the synthesis of a smooth adaptive control policy that achieves PTC of EL system trajectories subject to both state and input constraints while also undertaking a detailed feasibility analysis that illuminates the narrow operating region characterizing the interplay of SIT constraints. Thus, the main contributions of this study are the following: 

\begin{enumerate}
    \item {An approximation-free continuous adaptive barrier control policy is synthesized to guarantee local prescribed-time convergence of the tracking error for unknown EL systems subject to state and input constraints.}
    \item Full state constraint satisfaction is ensured by introducing a novel state constraint law, while input constraints are incorporated through a transformation technique using the saturation function. To the best of the author's knowledge, this is the only framework that imposes state and input constraints in achieving tracking error convergence in a prescribed-time to a prescribed-bound.
    \item Notably, detailed feasibility conditions are derived based on a set-theoretic approach that provides sufficient conditions for (i) the minimum control effort needed to drive the tracking errors to the prescribed-bound in a user-defined finite time and (ii) the upper bound on the external disturbances that the controller can accommodate during tracking.
    \item In addition, a viable set of initial conditions that guarantee convergence to the prescribed-bound set in the presence of prescribed SIT constraints is derived using higher-order control barrier functions.
\end{enumerate}

A detailed simulation study is provided for different robotic systems that highlight the efficacy of the proposed scheme. In addition, the salient features of the proposed controller are highlighted through a detailed qualitative and quantitative comparison with other leading alternative designs. 

The remainder of the paper is organized as follows. Section \ref{sec:problem_formulation} introduces the mathematical preliminaries and the control objectives. Section \ref{sec:feasibility_constraint} derives feasibility conditions for constrained systems. Section \ref{sec:adaptive_controller} presents the state constraint law and the adaptive barrier control policy followed by stability analysis. Section \ref{sec:results} provides a comprehensive performance analysis for two different industrial manipulators and a detailed comparison analysis with other related studies. Section \ref{sec:conclusion} presents the conclusion of this study and the scope for future work.

\section{Problem Formulation} \label{sec:problem_formulation}
This section introduces the preliminaries and the mathematical formalism illustrating the properties of EL systems and the control objectives of this study, followed by devising the tracking error with TBG for the control policy design.
\subsection{Preliminaries}
\noindent\emph{Notations}: We use bold letters to represent matrices and vectors. The sets of all real and nonnegative real numbers are denoted by $\R$ and $\R_{\geq0}$, respectively. Let $\N_n = \{1, 2, \cdots, n\}$. The vector inequalities, $\bs{a} \preceq \bs{b}\  (\bs{a} \succeq \bs{b})$ with $\bs{a},\ \bs{b} \in \R^n$ represents $a_i \leq b_i\ (a_i \geq b_i),\ \forall i \in \N_n$. $\bs{I}_n \in \R^{n \times n},\ \text{and}\ \bs{0}_n,\ \bs{1}_n \in \R^n$ represent the identity matrix and vector of zeros and ones, respectively. For square matrices $\bs{A}\ \text{and}\ \bs{B}$, $\bs{A} \leq \bs{B}$ states that $\bs{B} - \bs{A}$ is a positive semidefinite matrix. The $p$-norm, and multi-variable sign function are represented by $\norm{\cdot}$ and $\mSig{a} = \bs{a}/\norm{\bs{a}}$, respectively. $\lambda_{\Min}\{\cdot\}\ \text{and}\ \lambda_{\Max}\{\cdot\}$ represent the minimum and maximum eigenvalues of a given square matrix, respectively. Let $\text{abs}\paranthesis{\bs{a}} = [\abs{a_1},\ \dots,\ \abs{a_n}]^\top$.
{The subsequent notion of prescribed-time tracking allows the convergence time $0 < T < \infty$ and tracking a given reference point $\bs{x}_r\in\R^p$ with tracking error upper bound $\epsilon > 0$ to be chosen \emph{apriori} for a given system $\dot{\bs{x}} = \bs{f}(t, \bs{x}, \bs{u}, \bs{w})$ with $\bs{f}:\R_{\geq 0}\times\mathcal{X}\times\mathcal{U}\times\mathcal{W} \rightarrow \R^p$ where $\bs{x}\in\mathcal{X}\subset \R^p$, $\bs{u}\in\mathcal{U}\subset\R^m$, and $\bs{w}\in\mathcal{W}\subset\R^d$ represent the system's state, the control input and the external disturbance, respectively and $\mathcal{X}, \mathcal{U},\mathcal{W}$ are compact sets. Let $\bs{\Psi}_{\bs{\mathfrak{u}}}(t,t_0,\bs{x}(t_0)) \in \R^p$ represent the value of the trajectory of the system at time $t>t_0$ with $t_0 \geq 0$ and starting from the initial state $\bs{x}(t_0)$ under the input signal $\bs{\mathfrak{u}}$. Then, define a set ${\mathcal{B}_\epsilon(\bs{p})}$ as a ball of radius $\epsilon$, for any $\bs{p}\in\R^p$ as follows:}
\begin{equation}
    {\mathcal{B}_\epsilon(\bs{p})} = \{\bs{x} \in \R^p : \norm{\bs{x} - \bs{p}} \leq \epsilon\},
    \label{eq:ptcuel_prescribed_bound_set}
\end{equation}
and let $|\bs{a}|_{\mathcal{C}}=\inf_{\bs{b}\in{\mathcal{C}}}\norm{\bs{a}-\bs{b}}$ denote the distance of a point $\bs{a}$ from the set $\mathcal{C}\subset\R^p$.

\begin{definition}\label{def:ptpb_def}
{For user-defined constants $T,\epsilon > 0$, the equilibrium point $\bs{x}_r\in\mathcal{X}$ of the system $\dot{\bs{x}} = \bs{f}(t, \bs{x}, \bs{u}, \bs{w})$ with $\bs{x}\in\mathcal{X}$, $\bs{u}=\bs{\mathfrak{u}}(t,\bs{x})\in\mathcal{U}$, and $\bs{w}\equiv\bs{0}_q$ is called as \emph{local prescribed-time prescribed-bound} (PTPB) stable if the closed-loop trajectories of the system reach and remain in the ball $\mathcal{B}_\epsilon(\bs{x}_r)$ for any $\bs{x}(t_0)\in\mathcal{C}\subset\mathcal{X}$ within the prescribed time $T$, i.e. $|\bs{\Psi}_{\bs{\mathfrak{u}}}(t, t_0, \bs{x}(t_0))|_{\mathcal{B}_\epsilon(\bs{x}_r)} = 0$ $\forall\ t\geq t_0+T$, where $\mathcal{C}$ is some neighbourhood of $\bs{x}_r$.}
\end{definition}

\begin{definition}\label{def:robust_ptpb_def}
{For user-defined constants $T,\epsilon > 0$, the equilibrium point $\bs{x}_r\in\mathcal{X}$ of the system $\dot{\bs{x}} = \bs{f}(t, \bs{x}, \bs{u}, \bs{w})$ with $\bs{x}\in\mathcal{X}$, $\bs{u}=\bs{\mathfrak{u}}(t,\bs{x})\in\mathcal{U}$, and $\bs{w}\in\mathcal{W}$ is called as \emph{robust local prescribed-time prescribed-bound} (PTPB) stable if the closed-loop trajectories of the system reach and remain in the ball $\mathcal{B}_\epsilon(\bs{x}_r)$ for any $\bs{x}(t_0)\in\mathcal{C}\subset\mathcal{X}$ within the prescribed time $T$, i.e. $|\bs{\Psi}_{\bs{\mathfrak{u}}}(t, t_0, \bs{x}(t_0))|_{\mathcal{B}_\epsilon(\bs{x}_r)} = 0$ $\forall\ t\geq t_0+T$, where $\mathcal{C}$ is some neighbourhood of $\bs{x}_r$.}
\end{definition}

\begin{remark}
    {Note that \emph{Definition} \ref{def:ptpb_def} for local PTPB stability is different from the definition of uniform asymptotic stability (UAS) given in \cite[Definition 4.4]{khalil}. Given $c>0,\ \bs{x}_r = \bs{0}_p,\ \mathcal{C}=\mathcal{B}_c(\bs{x}_r)\subset\mathcal{X}$, and $\bs{u} = \bs{\mathfrak{u}}(t,\bs{x})\in\R^m$ verifies UAS, then according to the definition of UAS, the convergence time depends on the bound $\epsilon$ for a closed-loop system, that is, for each $\epsilon_1>0$ assume that $c>\epsilon_1 > \epsilon$, there is $T = T(\epsilon_1)>0$ such that $\norm{\bs{\Psi}_{\mathfrak{\bs{u}}}(t,t_0,\bs{x}(t_0)}<\epsilon_1\ \forall\ t \geq t_0 + T(\epsilon_1)\ \forall\ \bs{x}(t_0)\in\mathcal{C}$. Thus, convergence time depends on the prescribed bound $\epsilon$. On the other hand, local PTPB stability states that the convergence time $T$ is independent of the bound $\epsilon$, that is, $\norm{\bs{\Psi}_{\mathfrak{\bs{u}}}(t,t_0,\bs{x}(t_0)}<\epsilon,\ \forall t\geq t_0+T$.}
\end{remark}

\begin{remark}
  {The \emph{Definition} \ref{def:ptpb_def} for local PTPB stability with $\bs{u} = \bs{0}_m$ is more general than \cite[Definition 2]{song:2017} and \cite[Definition 1]{kgarg_clf}, which address the constraint-free case $\bs{x} \in \R^p, \bs{u}\in\R^m$, and $\mathcal{C} = \R^p$. Likewise, \emph{Definition} \ref{def:robust_ptpb_def} for robust local PTPB stability with $\bs{u} = \bs{0}_m$ is more general than \cite[Definition 1]{yeCao}, where PTPB stability is defined without accounting for external disturbances.}
\end{remark}

We define the following constraint sets for imposing the SIT constraints for a given control law $\bs{\mathfrak{u}}(t,\bs{x})$.
\begin{flalign}
    \C_1 =& \{\bs{x} \in \llbracket\bs{x}^-, \bs{x}^+\rrbracket : \bs{\mathfrak{u}}(t,\bs{x}) \in \llbracket\bs{u}^-, \bs{u}^+\rrbracket, \forall t\geq0 \} \label{eq:ptcuel_state_constraints_set},
\end{flalign}
\begin{flalign}
    \C_2 =& {\{ \bs{x}(t_0) \in \R^p: {\bs{u}\in\llbracket\bs{u}^-, \bs{u}^+\rrbracket}\ \text{and}\ \forall t\geq t_0+T,} \nonumber\\& {\bs{\Psi}_{\bs{\mathfrak{u}}}(t, t_0, \bs{x}(t_0))\in{\mathcal{B}_\epsilon(\bs{x}_r)} \}},
    \label{eq:ptcuel_viable_set}
\end{flalign}
where $\bs{x}^-,\ \bs{x}^+ \in \R^p$ and $\bs{u}^-,\ \bs{u}^+\in\R^m$ are bounds on state and inputs, respectively. Also, $\llbracket\bs{a},\bs{b}\rrbracket$ denote hyper-interval $[a_1, b_1]\times[a_2,b_2]\times \cdots \times[a_n,b_n]$.

Achieving PTC from arbitrarily large initial conditions in the presence of input constraints may not always be possible. Consequently, there exists a relation between prescribed-time $T$ and input constraints. In the following Lemma, the lower bound on $T$ is derived for the control affine system.

\Lemma{
    For a given control-affine system under input constraints, $\dot{\bs{x}} =\bs{f}_a(\bs{x}) + \bs{g}_a(\bs{x})\bs{u}$ with $\bs{f}_a:\R^p\rightarrow\R^p$ and $\bs{g}_a:\R^{p}\rightarrow\R^{p\times m}$, $\norm{\bs{u}} \leq u^*=\Min\paranthesis{\norm{\bs{u}^-},\norm{\bs{u}^+}}$, user-defined constant $\epsilon>0$ and a reference point $\bs{x}_r\in\R^p$. {Suppose there exist real positive constants $\ol{f},\ \ol{g}$ such that $\norm{\bs{f}_a(\bs x)}\leq \ol{f}$ and $\norm{\bs{g}_a(\bs x)}\leq \ol{g}$ for all $x\in\mathbb{R}^p$, and a continuous feedback control law $\bs{u} = \bs{\mathfrak{u}}(t,\bs{x})$, then prescribed-time $T$ is lower bounded} as 
    \begin{flalign}
         {T \geq T^*= \underset{\bs{x}(t_0)\in\mathcal{C}_2}{\sup}\left \{\frac{|\bs{x}(t_0)|_{\mathcal{S}_\epsilon(\bs{x}_r)}}{\ol{f}+\ol{g}u^*}\right \}},
        \label{eq:ptcuel_PT_lower_bound}
    \end{flalign}
    such that control law drives the solution $\bs{\Psi}_{\bs{\mathfrak{u}}}(t, t_0, \bs{x}(t_0))$ to the prescribed-bound set ${\mathcal{S}_\epsilon(\bs{x}_r)}$, i.e. {$\bs{\Psi}_{\bs{\mathfrak{u}}}(t, t_0, \bs{x}(t_0))\in{\mathcal{S}_\epsilon(\bs{x}_r)},\ \text{for all}\ t\geq t_0+T$, and $\bs{x}(t_0)\in\mathcal{C}_2$}}.

{\label{ptcuel_Lemma}}

For proof, see \emph{Appendix} \ref{Lemma_proof}. {Note that the controller in \emph{Lemma} \ref{ptcuel_Lemma} assumes that the controller $\bs{\mathfrak{u}}(t,\bs{x})$ guarantees local prescribed-time stability for all $\bs{x}(t) \in \mathcal{C}_2$ as per \emph{Definition} \ref{def:ptpb_def}}. Also, from \emph{Lemma} \ref{ptcuel_Lemma}, it is clear that if there are no input constraints, one can prescribe an arbitrarily small convergence time $T>0$ given an initial condition $\bs{x}(t_0)$. However, the presence of input constraints prevents the selection of an arbitrarily small value of the convergence time parameter $T$, so that a \emph{feasible} estimate for this parameter can be obtained based on the inequality \eqref{eq:ptcuel_PT_lower_bound}. {Consequently from \emph{Lemma} 1, for all states $\bs{x}(t) \in \mathcal{C}_2$ implies locally PTS.}

\Remark{{Note that the computation of the set $\mathcal{C}_2$ is significantly challenging in general. Moreover, this study focuses on EL systems subjected to SIT constraints. Hence, we provide a viable set of initial conditions {(i.e., $\Set$ in Fig. \ref{fig:ptcuel_set_theoretic_demo})} as in \eqref{eq:ptcuel_analytical_viable_set} using higher-order control barrier functions, which is presented in \emph{Theorem} \ref{ptcuel_feasibility_theorem}. }}{\label{rem:set_C2}}

\subsection{Problem Statement} \label{sec:ptcuel_problem_statement}
Consider the class of EL systems with $\q:\R_{\geq0}\rightarrow\R^n$ being position coordinates, $\qd,\ \qdd$ denote the first and second order derivatives of position coordinates, respectively, defined as below: 
\begin{equation}
    \bs{M}(\q)\qdd + \bs{C}(\q, \qd)\qd + \bs{G}(\q) + \bs{F}(\qd) + \boldsymbol{d}(t) = \bs{u},
    \label{eq:ptcuel_eom}
\end{equation}
where $\bs{M}(\q) \in \R^{n \times n}$ is mass matrix, $\bs{C}(\q, \qd) \in \R^{n \times n}$ is Coriolis matrix, $\bs{G}(\q) \in \R^n$ is gravity vector, $\bs{F}(\qd) \in \R^n$ is damping and friction forces, $\bs{d}(t) \in \R^n$ is external disturbance and $\bs{u} \in \R^n$ is the control input. For brevity, when a symbol’s functional dependence is clear, its arguments and brackets are omitted; e.g., $\bs{C}(\q,\qd)$ and $\bs{d}(t)$ are written as $\bs{C}$ and $\bs{d}$, respectively.

Further, the properties of EL systems \cite[Chapter 2]{EL_props},  \cite{Robot_Control:1990} are stated below for some positive real constants $\udl{M},\ \ol{M},\ \udl{m},\ \ol{m},\ \ol{C},\ \ol{G},\ \ol{F}$ that represent bounds on norm of the system matrices.

\prop{The matrix $\dot{\bs{M}}-2\bs{C}$ is skew-symmetric.} {\label{prop:skew}}

\prop{$\udl{M}\bs{I}_n{\leq}\bs{M} {\leq} \ol{M}\bs{I}_n$ and $\udl{m}\bs{I}_n {\leq} \bs{M}^{-1} {\leq} \ol{m}\bs{I}_n$.} {\label{prop:mass}}

\prop {$\norm{\bs{C}}\leq \ol{C}\norm{\qd}$, $\norm{\bs{G}}\leq \ol{G}$, $\norm{\bs{F}}\leq \ol{F}\norm{\qd}$. } {\label{prop:matrix_bounds}}

Let $\bs{{u}}^+\succ\bs{0}_n,\ \bs{u}^-\prec\bs{0}_n,\ \bs{\theta}_q^-,\ \bs{\theta}_q^+,\ \bs{\nu}_q^-,\ \text{and}\ \bs{\nu}_q^+ \in \R^n$ be some known constants such that state and input constraints for the system \eqref{eq:ptcuel_eom} is defined for all $t \in \R_{\geq0}$. Also, let state $\bs{x} {=} [\q^\top, \qd^\top]^\top$, $\bs{x}^-{=} [(\bs{\theta}_q^-)^\top, (\bs{\nu}_q^-)^\top]^\top$, $ \bs{x}^+ {=} [(\bs{\theta}_q^+)^\top, (\bs{\nu}_q^+)^\top]^\top$. Then following sets defined as: state constraints $\mathcal{X} = \llbracket\bs{x}^-,\bs{x}^+\rrbracket$, and input constraint set $\mathcal{U} = \llbracket\bs{u}^-,\bs{u}^+\rrbracket$ for the system in \eqref{eq:ptcuel_eom}. Before formally stating the control objective, the following assumptions are invoked in this study:

\assume{The system matrices $\bs{M},\ \bs{C},\ \bs{G},\ \bs{F},\ \text{and}\ \bs{d}$ are assumed to be unknown for deriving the control law.} {\label{assume:system_matrices}}

\assume {The external disturbance $\bs{d}(t)$ is bounded by some unknown positive  constant, $\ol{d}$, i.e. $\norm{\bs{d}(t)}\leq \ol{d},\ \forall\ t\geq 0$.} {\label{assume:bounded_extn}}

\assume{For a given reference trajectory $\dq(t),\ \dqd(t),\ \text{and}\ \dqdd(t)$, there exist known constants $\ol{q}_r,\ \ol{q}_{rd},\ \text{and}\ \ol{q}_{rdd}$ such that $\norm{\dq} \leq \ol{q}_r,\ \norm{\dqd} \leq \ol{q}_{rd},\ \text{and}\ \norm{\dqdd} \leq \ol{q}_{rdd}$. In addition, the reference trajectory satisfies the following inequality:
\begin{equation}
    \bs{\theta}_q^- \preceq \dq(t) \preceq \bs{\theta}_q^+,\  \bs{\nu}_q^- \preceq \dqd(t) \preceq \bs{\nu}_q^+,\,\forall\,t\geq 0.
    \label{eq:ptcuel_des_state_condition}
\end{equation}
} {\label{assume:traj_bound}}
A practical constraint requires system \ref{eq:ptcuel_eom} to have sufficient control authority to drive the states to the reference signals $\boldsymbol{q}_r(t)$ and $\dot{\boldsymbol{q}}_r(t)$.
\begin{figure}[!t]
    \centering
    \includegraphics[width=0.25\linewidth]{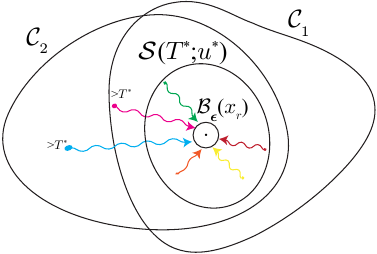}
    \caption{{Illustrating interplay of SIT constraints on the system.}}
    \label{fig:ptcuel_set_theoretic_demo}
\end{figure}
Note that \emph{Properties} \ref{prop:mass}, \ref{prop:matrix_bounds}, and \emph{Assumption} \ref{assume:system_matrices} are not used in deriving the control law (Section \ref{sec:adaptive_controller}). Unlike prior works \cite{PTC_TBG, TBG_Higher_order}, which assume bounded disturbances and their derivatives, this study (\assumeref{\ref{assume:bounded_extn}}) only assumes bounded disturbances. \emph{Assumptions} {\ref{assume:traj_bound}} are practically motivated.

{To this end, by the concurrent imposition of state and input constraints, the region of operation is reduced to $\C_1$. Moreover, with the inclusion of input and temporal constraints as in \emph{Lemma} 1, this region of operation is further lessened as $\C_1\cap\C_2$}. As illustrated in Fig. \ref{fig:ptcuel_set_theoretic_demo}, the goal is to drive the system state $\bs{x} \in \C_1\cap\C_2$ to the set ${\mathcal{B}_\epsilon(\bs{x}_r)}$ within the prescribed-time $T$ {to achieve local PTS}. With this objective, let ${\Set} \subset \R^{2n}$ denote the set of initial conditions {with prescribed-time $T$ and input constraints $u^*$ as in \emph{Lemma} \ref{ptcuel_Lemma}}, the problem statement is formally stated as below:

\prob{Given an Euler-Lagrangian system \eqref{eq:ptcuel_eom} satisfying \textit{Assumptions} \ref{assume:system_matrices} -- \ref{assume:traj_bound} subjected to state constraints ($\bs{x}\in\mathcal{X}$) and input constraints ($\bs{u}\in\mathcal{U}$), {the control objective is to synthesize a model parameter-free feedback control strategy $\bs{u}$ that achieves robust local PTPB convergence (as in \emph{Definition} \ref{def:robust_ptpb_def}) for any $\bs{x}_0 \in \Set$ in a user-defined prescribed time $T$ satisfying \emph{Lemma} \ref{ptcuel_Lemma}.}}{\label{problem_statement}}

\subsection{Transformed Tracking Errors with TBG} \label{sec:ptcuel_tracking_error}

In this subsection, we devise a filtered tracking error with the aid of TBG, which we use to develop a controller that can adhere to SIT constraints.

To begin, the tracking errors are defined as $\err, \ed: \R_{\geq0}\times\R^n\rightarrow\R^n$ such that
\begin{equation}
    \err(t, \q(t)) = \q(t) - \dq(t),\, \ed(t,\qd(t)) = \qd(t) - \dqd(t).
    \label{eq:ptcuel_tracking_errors}
\end{equation}
In addition, for the states $\bs{x}\in\mathcal{X}$ can be used to show that $\forall\ t\in \R_{\geq0},\ \norm{\q(t)} \in [{\theta}^-_q,\, {\theta}^+_q]\ \text{and} \ \norm{\qd(t)} \in [{\nu}^-_{q},\, {\nu}^+_{q}] $, where ${\theta}^+_q = \norm{\bs{\theta}_q^+},\, {\theta}^-_q = \norm{\bs{\theta}_q^-},\, {\nu}^+_q = \norm{\bs{\nu}_q^+},\ \text{and}\ {\nu}^-_q=\norm{\bs{\nu}_q^-}$. It then follows that $\err^-(t) \preceq \err(t,\q) \preceq  \err^+(t) \ \text{and}\ \ed^-(t) \preceq \ed(t,\q) \preceq \ed^+(t)$, where $\err^{-}(t) = \bs{\theta}_q^{-} - \dq(t),\ \err^{+}(t) = \bs{\theta}_q^{+} - \dq(t),\ \ed^{-}(t) = \bs{\nu}_q^{-} - \dqd(t),\ \text{and}\ \ed^{+}(t) = \bs{\nu}_q^{+} - \dqd(t)$. Moreover, one can conclude that dynamic bounds \cite{ZHANG2008475} can be used to constrain $\ed(t,\qd)$ as
\begin{equation}
    \Tilde{\bs{e}}^- \preceq \ed \preceq  \Tilde{\bs{e}}^+,
    \label{eq:ptcuel_dynamicStateBounds}
\end{equation}
where $\Tilde{\bs{e}}^-(t) = \Max\{\ed^-(t), \kappa(\err^-(t) - \err(t,\q)) \},\ \Tilde{\bs{e}}^+(t) = \Min\{\ed^+(t), \kappa(\err^+(t) - \err(t,\q)) \}\ \text{and}\ \kappa>0$ is chosen to satisfy $\kappa > \underset{i \in \N_n}{\Max} \left \{\frac{\nu^+_{q,i} - \nu^-_{q,i}}{{\theta}^+_{q,i} -\theta^-_{q,i}} \right\}$, with $\theta^-_{q,i},\ \theta^+_{q,i},\  \nu^{-}_{q,i},\ \text{and}\ \nu^{+}_{q,i}$ are components of $\bs{\theta}_q^{-},\ \bs{\theta}_q^+,\ \bs{\nu}_q^-,\ \text{and}\ \bs{\nu}_q^{+}$, respectively.

For convergence of tracking errors $\err,\, \ed$ in prescribed-time $T$, the following TBG functions are employed. Let $h_j(t)$ be a fifth-degree polynomial subjected to the constraints $h_1(t_0) {=} \dot{h}_2(t_0) {=} 1,\, h_2(t_0) {=} \dot{h}_1(t_0) {=} \ddot{h}_j(t_0) {=} 0$ and $h_j(t {\geq} t_0{+}T) {=} \dot{h}_j(t {\geq} t_0{+}T) {=} \ddot{h}_j(t{\geq} t_0{+}T) {=} 0,\, \forall j {=} 1, 2$. Then, the coefficients of the fifth-degree polynomial are computed similarly to \cite{PTC_TBG}, and the resultant expressions for $h_1(t)$ and $h_2(t)$ are given as follows:
\begin{equation}
    h_1(t){ = }\!
    \begin{cases}
     {-}{6}(t{-}t_0)^5/T^5 {+} {15}(t{-}t_0)^4/T^4 &  \\
     {- }10(t{-}t_0)^3/T^3 {+} 1,& t {\in} [t_0, t_0{+}T)\\
    0,& t {\in} [t_0{+}T, \infty),
    \end{cases}
    \label{eq:ptcuel_tbg_h1}
\end{equation}
\begin{equation}
    h_2(t) {=} \!
    \begin{cases}
     {-}{3}(t{-}t_0)^5/{T^4} {+} {8}(t{-}t_0)^4/{T^3} & \\
     {-} {6}(t{-}t_0)^3/{T^2} {+} t, & t {\in} [t_0, t_0{+}T)\\
    0 ,& t {\in} [t_0{+}T, \infty).
    \end{cases}
    \label{eq:ptcuel_tbg_h2}
\end{equation}
Following this, we transform the tracking errors using TBG with $\eps, \epsd: \R_{\geq0}\times\R^n\rightarrow\R^n$ as follows:
\begin{flalign}
    \bs{\varepsilon}(t, \q) &= \err(t,\q) - \err^d(t),\ \dot{\bs{\varepsilon}}(t,\qd) = \ed(t,\qd) - \ed^d(t),\label{eq:ptcuel_tracking_error} \\
    \err^d(t) &{=} h_1(t)\err_0 {+} h_2(t)\err_{d_0},\ \ed^d(t) {=} \dot{h}_1(t)\err_0 {+} \dot{h}_2(t)\err_{d_0}, \label{eq:ptcuel_error_refs}
\end{flalign}
where $\err_0=\err(t_0),\, \err_{d_0} = \ed(t_0)$ and observe that in \eqref{eq:ptcuel_error_refs}, the functions $\err^d(t),\, \ed^d(t),\, \bs{\ddot{\err}}^d(t)$ are bounded by virtue of the functions $h_1(t),\, h_2(t)$, and let $\norm{\err^d(t)} \leq \ol{e},\, \norm{\ed^d(t)}\leq \ol{e}_{d},\, \norm{\bs{\ddot{\err}}^d(t)} \leq \ol{e}_{dd}$, where  
\begin{flalign}
    \ol{e} &= \norm{\err_0}+\frac{k_1}{T}\norm{\err_{d_0}},\ \ol{e}_d = \frac{k_2}{T}\norm{\err_0}+\norm{\err_{d_0}}, \nonumber\\ 
    \ol{e}_{dd} &= \frac{k_3}{T^2} \norm{\err_0}+\frac{k_4}{T}\norm{\err_{d_0}},
    \label{eq:ptcuel_tbg_bounds}
\end{flalign}
where $k_1 = 2,\ k_2 = 15/8,\ k_3 = 10\sqrt{3}/3,\ k_4 = (152\sqrt{19}+224)/225$. Following this, we provide feasibility analysis for EL system \eqref{eq:ptcuel_eom} in \ref{sec:feasibility_constraint} and design an adaptive control strategy along with stability analysis based on the transformed tracking errors developed in section \ref{sec:adaptive_controller}.

\section{Feasibility Analysis}
\label{sec:feasibility_constraint}
This section provides a feasibility analysis based on set-theoretic methods using a constraint set formed by higher-order control barrier functions (CBF).

To this end, note that \emph{Lemma} \ref{ptcuel_Lemma} dictates that there exists a viable set of initial conditions {under the influence of input constraints. However, an arbitrarily small settling time cannot guarantee the state constraint satisfaction as illustrated in Fig. \ref{fig:ptcuel_set_theoretic_demo}, since a shorter time to reach a given target position from any initial conditions implies a larger rate of change of position, i.e. a larger velocity bound. Thus, full state constraints on both position and velocity preclude the selection of an arbitrarily small convergence time. Consequently, there exists a domain of operation under the simultaneous involvement of state, and input constraints for PTPB convergence. Thus, a viable set of initial conditions, ${\Set}$ is derived, i.e. from any point within the set ${\Set}$, the state $\bs{x}(t)$ can reach ${\mathcal{S}_\epsilon(\bs{x}_r)}$ in the presence of SIT constraints to verify local PTPB convergence.} In this regard, we invoke forward invariance properties \cite{CBF_SC, setTheoretic} of the constrained set for EL systems to derive the set ${\Set}$. In addition, for a given state belonging to ${\Set}$, prescribed-time $T$, and state constraints, we also derive feasible conditions for sufficient control authority  and the maximum disturbance that the controller can handle.

Now consider the functions, $\ol{\bs{\zeta}}, \udl{\bs{\zeta}}:\R^n\rightarrow\R^n$ and construct the constraint set $\C_3$ as:
\begin{flalign}
    &\C_3 = \{\bs{p}\in\R^n: \ol{\bs{\zeta}}(\bs{p}) \succeq \bs{0}_n, \udl{\bs{\zeta}}(\bs{p})\succeq\bs{0}_n\} \label{eq:ptcuel_first_order_set},\\
    &\ol{\bs{\zeta}}(\bs{p}) = \bs{\theta}_q^+ - \bs{p},\, \udl{\bs{\zeta}}(\bs{p})=\bs{p} - \bs{\theta}_q^-.\label{eq:ptcuel_first_order}
\end{flalign}
Then, the forward invariance properties of the set, as required by Nagumo's theorem, can then be computed as:
\begin{equation}
    \dot{\ol{\bs{\zeta}}}(\q) \succeq -\bs{\alpha}(\ol{\bs{\zeta}}(\q)),\, \dot{\udl{\bs{\zeta}}}(\q) \succeq -\bs{\alpha}(\udl{\bs{\zeta}}(\q)),
    \label{eq:ptcuel_fwd_first_order}
\end{equation}
where $\bs{\alpha}(\bs{a}) = [\alpha_1(a_1),\ \cdots,\ \alpha_n(a_n)]^\top$, $\alpha_i(a_i)$ is an extended class $\mathcal{K}$ function \cite{CBF_SC} and $a_i$ is a component of the vector $\bs{a}$ for all $i \in \N_n$. However, the system \eqref{eq:ptcuel_eom} is of relative degree 2, therefore no control input exists that can ensure the forward invariance properties of the set, $\C_3$. So, we define the higher-order CBF \cite{CBF_EL_IP} with $\ol{\bs{\beta}}(t,\bs{p},\bs{v}), \udl{\bs{\beta}}(t,\bs{p},\bs{v}):\R_{\geq0}\times\R^n\times\R^n \rightarrow\R^n$ as follows:
\begin{flalign}
    \C_4 &= \{(\bs{p},\bs{v})\in\R^{2n}: \ol{\bs{\beta}} \succeq \bs{0}_n,\ \udl{\bs{\beta}} \succeq \bs{0}_n\}, \label{eq:ptcuel_higher_order_set}\\
    \ol{\bs{\beta}} &= -\epsd + \sigma\bs{\alpha}(\ol{\bs{\zeta}}),\ \udl{\bs{\beta}}  =\epsd + \sigma\bs{\alpha}(\udl{\bs{\zeta}}) ,\label{eq:ptcuel_higher_order}
\end{flalign}
where $\bs{p}$ and $\bs{v}$ represents position and velocity coordinate, respectively. Also, $\sigma>0$ is some constant that has to be chosen appropriately for a given system \eqref{eq:ptcuel_eom} to render the viable set (see \emph{Theorem} \ref{ptcuel_feasibility_theorem}). Note that $\epsd$ is  bounded by virtue of the state $\qd$ as seen from \eqref{eq:ptcuel_dynamicStateBounds}, \eqref{eq:ptcuel_error_refs} and \eqref{eq:ptcuel_higher_order_set}. By setting $\alpha_i(a_i)=a_i,\ \bs{p} = \q(t),\ \bs{v} = \qd(t),\ \text{and}\ \dq(t) = \q^*$, where $\q^*$ as a constant reference point, the lower bound on $\sigma$ is computed as follows:
\begin{flalign}
    -\sigma\paranthesis{\q - \bs{\theta}_q^-} \leq \epsd \leq \sigma\paranthesis{\bs{\theta}_q^+ - \q }.\label{eq:ptcuel_eps_bounds}
\end{flalign}
Now, taking the maximum of a lower bound and upper bound on $\epsd$ yields the following:
\begin{flalign}
    \norm{\epsd}_\infty &\leq \Max \left \{ \underset{\forall i \in \N_n}{\Max}\sigma \paranthesis{\theta_{q,i}^+{-}q_i},\; \underset{\forall i \in \N_n}{\Max}\sigma \paranthesis{q_i{-}\theta_{q,i}^-}\right \}, \nonumber \\
    \norm{\qd-\ed^d}_\infty &\leq \sigma\norm{\bs{\theta}_q^+-\bs{\theta}_q^-}_\infty {\implies}
    \sigma \geq \frac{\norm{\qd - \ed^d}_\infty}{\norm{\bs{\theta}_q^+-\bs{\theta}_q^-}_\infty}, \label{eq:ptcuel_gamma_lower_1}
\end{flalign}
with $q_i,\ \theta_{q,i} ^-, \text{and}\ \theta_{q,i}^+$ are components of $\q,\ \bs{\theta}_q^-, \text{and}\ \bs{\theta}_q^+,\ \forall\ i \in \N_n$, respectively. In order to obtain the tighter bound on the inequality \eqref{eq:ptcuel_gamma_lower_1}, consider the $\Max\{\norm{\qd - \ed^d}_\infty\}$ as follows:
\begin{flalign}
    \norm{\qd- \ed^d}_\infty \leq \norm{\qd}_\infty + \norm{\ed^d}_\infty 
    &\leq \nu^+_q {+} \ol{e}_{d,\Max} \nonumber \\
    &\leq 2\Max\paranthesis{\nu^+_q, \ol{e}_{d,\Max}}, \label{eq:ptcuel_gamma_strange_upper}    
\end{flalign}
\begin{equation}
    \ol{e}_{d,\Max} = \underset{\bs{x}(t_0){\in} \C_1}{\Max}\paranthesis{\frac{k_2}{T}\norm{\q(t_0){-}\q^*}_\infty {+}  \norm{\qd(t_0)}_\infty}. \label{eq:ptcuel_e_d_max_upper}
\end{equation}
Then, the lower bound can be computed as $\sigma \geq \udl{\sigma} =  2\Max\paranthesis{\nu^+_q, \ol{e}_{d,\Max}}/\norm{\bs{\theta}_q^+-\bs{\theta}_q^-}_\infty$. To present the next \emph{Theorem} \ref{ptcuel_feasibility_theorem}, we define the following terms:
\begin{flalign}
    \ol{\sigma} &:= \frac{1}{\ol{M}\nu^+_q}\paranthesis{u^* - \paranthesis{\ol{C}(\nu_q^+)^2 + \ol{F}\nu_q^+ + \ol{G}}}, \label{eq:ptcuel_sigma_bar}\\
    \ol{f} &= \paranthesis{(\nu^+_q)^p {+} \paranthesis{\ol{m}\paranthesis{\paranthesis{\ol{C}\nu^+_q{+}\ol{F}}\nu^+_q{+}\ol{G}}}^p}^{\frac{1}{p}},\ \ol{g} = \ol{m}. \label{eq:ptcuel_fBar_gBar}
\end{flalign}
Note that the values of $\ol{f}$ and $\ol{g}$ in \eqref{eq:ptcuel_fBar_gBar} are computed for EL system \eqref{eq:ptcuel_eom} excluding the disturbance term $\bs{d}$.

In order to arrive at the viable set and obtain the lower bound on the sufficient control authority and maximum disturbances that the controller can handle to achieve PTPB convergence from a range of initial conditions to user-defined targets, we consider the constant reference target and propose the following theorem.
\begin{theorem} \label{ptcuel_feasibility_theorem}
Consider the system \eqref{eq:ptcuel_eom} with \textit{Properties} \ref{prop:skew}--\ref{prop:matrix_bounds} and under \textit{Assumptions} \ref{assume:system_matrices} -- \ref{assume:traj_bound}. For a given $\sigma \in [\udl{\sigma}, \ol{\sigma})$ and { for a sufficiently large prescribed-time $T$ chosen as in \emph{Lemma} \ref{ptcuel_Lemma}}, the viable set of initial conditions $\Set \subset \C_1$ is given as:
\begin{flalign}
    \Set {=} {\left \{{(\bs{q}, \dot{\bs{q}}}){\in}{\C_4} {:} \frac{k_3}{T^2}{\norm{\bs{q} {-} \q^*}} {+} \frac{k_4}{T} \norm{\dot{\bs{q}}} {\leq} {\frac{1}{\ol{M}}}{\paranthesis{{u^*} {-}{\eta}}} \right \}} ,
    \label{eq:ptcuel_analytical_viable_set}
\end{flalign}
where the set $\C_4 = \{(\bs{q}, \dot{\bs{q}})\in (\llbracket\bs{\theta}_q^-, \bs{\theta}_q^+ \rrbracket \times \R^n) \}\cap \C_3$, the constant $\q^*\in\llbracket\bs{\theta}_q^-, \bs{\theta}_q^+ \rrbracket$ is a reference point, $\bs{x}_r = [(\bs{q}^*)^\top, \bs{0}_n^\top]^\top$, ${u}^* = \Min\paranthesis{\norm{\bs{u}^-}, \norm{\bs{u}^+}}$, and $\eta = \ol{C}(\nu_q^+)^2+\ol{F}\nu_q^++\sigma \ol{M}\nu_q^+ + \ol{G}$ with $\ol{C},\ \ol{F},\ \ol{G},\ \ol{M}$ are bounds on system matrices (refer \emph{Properties} \ref{prop:mass} and \ref{prop:matrix_bounds}). Further, the lower bound on sufficient control authority $u_\Min$, and upper bound on the maximum disturbances $\ol{d}$, that the controller can handle are estimated such that $u^* \geq u_\Min$ as follows:
\begin{flalign}
    u_\Min & \geq \eta + \ol{M}\Max(\ol{e}_{dd}),\label{eq:ptcuel_min_control_authority}\\
    \ol{d} &\leq u^* -  \paranthesis{ \eta + \ol{M}\Max(\ol{e}_{dd})},\label{eq:ptcuel_max_disturbances} \\
    \Max(\ol{e}_{dd}) &{=} {\underset{\bs{x}(t_0){\in}{\Set}}{\Max}}\!\paranthesis{{\frac{k_3}{T^2}} {\norm{\q(t_0){-}\q^*}}{+} {\frac{k_4}{T}} {\norm{\qd(t_0)}}}. \label{eq:ptcuel_edd_max}
\end{flalign}
\end{theorem}

For proof, see the \emph{Appendix} \ref{ptcuel_feasibility_theorem_proof}. {Considering any given initial condition within the set $\Set$, one can still guarantee local PTPB convergence with a plausible control scheme}. Also, it is important to note that Theorem 1 allows a straightforward verification of the fact that $\C_5$ is forward invariant after the application of an appropriate control policy (designed in the later section) that guarantees the satisfaction of SIT constraints. In particular, we now undertake the design of an adaptive barrier function-based control policy that guarantees that the set $\C_5$ is forward invariant for a fixed $T$ and $u^*$.

\Remark{From \eqref{eq:ptcuel_min_control_authority}, it is clear that greater control effort is required to achieve tracking convergence for a lower prescribed convergence time $T$, and/or larger values of the initial state $\bs{x}(t_0)$. Additionally, from \eqref{eq:ptcuel_max_disturbances}, it is clear that the proposed controller can handle larger disturbances for larger values of the user-defined prescribed-time $T$ while retaining sufficient control authority to drive the tracking error to the prescribed-bound $\epsilon$ in the prescribed-time $T$. Furthermore, from \eqref{eq:ptcuel_analytical_viable_set}, it is evident that a larger feasible set ${\Set}$ may be realized for larger values of the prescribed-time $T$ and the input constraint $u^*$.}{\label{rem:feasibility}}

\Remark{\emph{Theorem} \ref{ptcuel_feasibility_theorem} may be used to arrive at a viable set of initial conditions \eqref{eq:ptcuel_analytical_viable_set} for an EL system \eqref{eq:ptcuel_eom} subjected to more restrictive SIT constraints, whereas the study \cite{kgarg_clf} derives a viable set when only subjected to input and temporal constraints. In addition, the proposed scheme also derives sufficient conditions for minimum control authority \eqref{eq:ptcuel_min_control_authority} and maximum disturbance rejection capability \eqref{eq:ptcuel_max_disturbances} for the controller, whilst the study \cite{CBF_EL_IP} only provides sufficient control authority conditions for an EL system \eqref{eq:ptcuel_eom} in the presence of state and input constraints. Thus, \emph{Theorem} \ref{ptcuel_feasibility_theorem} helps realize \emph{apriori} the domain of operation for successful tracking under the influence of SIT constraints.}{\label{rem:feasibility_comparison}}

\section{Adaptive Barrier Function Control Design} \label{sec:adaptive_controller}

This section establishes the state constraint equation using filtered tracking errors developed in \ref{sec:ptcuel_tracking_error} and presents the design of the adaptive barrier-function control law that incorporates this state constraint followed by stability analysis. 

\subsection{Adaptive Controller Design}
This subsection formulates the state constraint equation followed by constraining the control input through a non-linear transformation using a saturation function. Subsequently, an adaptive barrier control law based on an unconstrained variable is proposed to prevent the violation of state and input constraints. 

The filtered tracking error $\bs{\chi}: \R_{\geq0}\times\R^{2n} \rightarrow \R^n$ with $\bs{x} = [\q^\top, \qd^\top]$ is defined as:
\begin{equation}
    \bs{\chi}(t, \bs{x}(t)) = \epsd(t, \q(t)) + \bs{K}_p \eps(t,\qd(t)),
    \label{eq:ptcuel_filteredError}
\end{equation}
where $\bs{K}_p \in \R^{n \times n}$ is a user-defined positive-definite diagonal gain matrix.  Also, observe that $\err_0=\err^d(t_0)$ and $\err_{d_0}=\ed^d(t_0)$, thus $\bs{\chi}(t_0, \bs{x}(t_0))=\bs{0}_n$. Let $K_{p_{\Min}} = \underset{ i \in \N_n}{\Min}\{K_{p_i}\}\ \text{and}\ K_{p_{\Max}} =  \underset{ i \in \N_n}{\Max}\{K_{p_i}\}$ with $K_{p_i}$ as diagonal elements of matrix $\bs{K}_p$.

From \eqref{eq:ptcuel_dynamicStateBounds} and \eqref{eq:ptcuel_filteredError}, notice that $\bs{\chi}$ is bounded as $\bs{\varphi}^-_0(t) \preceq \bs{\chi}(t,\bs{x}) \preceq \bs{\varphi}^+_0(t)$ such that
\begin{equation}
    \bs{\varphi}^+_0(t) = \dot{\bs{\varepsilon}}^+(t) + \bs{K}_p\bs{\varepsilon},\,\bs{\varphi}^-_0(t) = \dot{\bs{\varepsilon}}^-(t) + \bs{K}_p\bs{\varepsilon},
    \label{eq:ptcuel_phi_minus}
\end{equation}
where $\dot{\bs{\varepsilon}}^-(t) = \Tilde{\err}^-(t) - \ed^d(t)$ and $\dot{\bs{\varepsilon}}^+(t) = \Tilde{\err}^+(t) - \ed^d(t)$. Consequently, by introducing $\bs{D} = [-\bs{I}_n,\ \bs{I}_n]^\top\in\R^{2n\times n}$, one can obtain the following inequalities:
\begin{equation}
    \bs{D}\bs{\chi}(t, \bs{x}) \preceq \bigl[-(\bs{\varphi}^{-}_0(t))^\top,\,(\bs{\varphi}^{+}_0(t))^\top\bigr]^{\top} = \bs{\Phi}_0(t).
    \label{eq:ptcuel_DynamicStateBoundsTBGCombined}
\end{equation}
Then, by introducing vector of positive definite real-valued functions  $\bs{y}(t) = \begin{bmatrix} y^2_1(t), & \cdots, & y^2_{2n}(t)\end{bmatrix}^\top$, with the unknown variable $y_i(t),\,i\in\N_{2n}$, one can reduce the inequality in \eqref{eq:ptcuel_DynamicStateBoundsTBGCombined} to a  equality as given in \eqref{eq:ptcuel_stateConstraint_equality}.
\begin{equation}
 \bs{D}\bs{\chi} + \bs{y} = \bs{\Phi}_0.
 \label{eq:ptcuel_stateConstraint_equality}
\end{equation}
Then, by introducing the new variable $\bs{\xi}:\R_{\geq0}\times\R^{2n}\rightarrow \R^{2n}$, and using \eqref{eq:ptcuel_stateConstraint_equality}, the state constraint can be reformulated as
\begin{equation}
    \bs{\xi}(t, \bs{x}) = \bs{D}\bs{\chi}(t, \bs{x}) + \bs{y}(t) - \bs{\Phi}(t),
    \label{eq:ptcuel_StateConstrainEquation}
\end{equation}
 where $\bs{\Phi}(t) = [-(\bs{\varphi}^-(t))^\top\ (\bs{\varphi}^+(t))^\top]^\top$, $\bs{\varphi}^-(t) = \bs{\varphi}^-_0(t) + c\bs{1}_n,\,\ \bs{\varphi}^+(t) = \bs{\varphi}^+_0(t) -c\bs{1}_n$ and $c>0$ denotes a user-defined safety margin. In the next section, it is shown that this safety margin can be appropriately chosen \emph{apriori} based on the ultimate bound of  $\bs{\xi}$ so that the control policy (to be designed as in \eqref{eq:ptcuel_Yupdate}) ensures rigorous satisfaction of state constraints while fulfilling the tracking objective.

Now, for incorporating the input constraints $\bs{u}\in\mathcal{U}$, the unconstrained input variable $\bs{\tau}(t) \in \R^n$ is introduced, which relates $\bs{u}(t)$ as $\bs{u}(t) = \bs{\Pi}(\bs{\tau}(t))\bs{\tau}(t)$, where $\bs{\Pi}(\bs{\tau}(t)) = \text{diag}\paranthesis{[\Pi_1(\tau_1)\ ,\cdots,\ \Pi_n(\tau_n)]} \in \R^{n \times n}$, is the diagonal matrix with diagonal entries given by
\begin{equation}
    {\Pi}_i({\tau}_i(t)) = \begin{cases}
        {u}^+_i/\tau_i, & \tau_i > u^+_i \\
        1, & u^-_i\leq\tau_i\leq u^+_i\\
        {u}^-_i/\tau_i, & \tau_i < u^-_i.
    \end{cases}
    \label{eq:ptcuel_SaturationFunction}
\end{equation}  
Then, the control objective translates to synthesizing the unconstrained input signal $\bs{\tau}(t)$ to drive the tracking errors to smaller bounds within user-defined time $T$. Inspired from \cite{barrierFunc}, we propose a barrier function-based adaptive controller accounting for the state-input constraints as in \policy{}.
\begin{equation}
    \begin{aligned}
        \bs{\tau}(t) &= -K(t)\Gamma(\bs{\varepsilon},  \dot{\bs{\varepsilon}})\mSig{\bs{\chi}} -  K(t)\norm{\bs{\xi}}\norm{\bs{D}}\mSig{\bs{\chi}},\\
        K(t) &= \frac{\varrho\norm{\bs{\chi}}}{\varpi - \norm{\bs{\chi}}},\ \\
        \Gamma(\bs{\varepsilon}, \dot{\bs{\varepsilon}}) &= 4\Max\{1,\  \norm{\bs{\varepsilon}(t)},\ \norm{\dot{\bs\varepsilon}(t)},\  \norm{\bs{\varepsilon}(t)} \norm{\dot{\bs\varepsilon}(t)}\}, \\
    \end{aligned}
    \label{eq:ptcuel_ControlPolicy}
\end{equation}
where $\varrho,\ \varpi >0$ are user-defined gains. Further, the computation of unknown functions $y_i(t),\ \forall\ i \in \N_{2n}$ is processed through the introduction of the variable $\bs{\Upsilon}(t) := \begin{bmatrix} y_1(t), & \cdots, & y_{2n}(t)\end{bmatrix}^\top \in \mathbb{R}^{2n}$ as
\begin{equation}
    \dot{\bs{\Upsilon}} = \frac{1}{2}\bs{\Lambda}^{-1}( -\gamma \bs{\xi} + \gamma\alpha \bs{D}\bs{\chi} + \dot{\bs{\Phi}} - \bs{D}\dot{\bs{\chi}}),
    \label{eq:ptcuel_Yupdate}
\end{equation}
where $\gamma>0$ and $0<\alpha\leq||\bs{D}||^{-1}$ are user-defined constants, $\bs{\Lambda}(t) = \text{diag}\paranthesis{\begin{bmatrix}y_1(t), & \cdots, & y_{2n}(t)\end{bmatrix} }$ and $\bs{\Upsilon}(t_0)$ is obtained from $\bs{y}(t_0)$ in \eqref{eq:ptcuel_StateConstrainEquation} by choosing $\bs{\xi}(t_0) = \bs{0}_n$ as $\bs{y}(t_0)=\bs{\Phi}(t_0)$.

The objective of the proposed strategy is to ensure that filtered tracking error, $\bs{\chi}$, lies within the ball of radius $\varpi$, while accommodating the state constraints through the variable $\bs{\xi}$.

\Remark{It is important to note that the adaptive control policy \policy{} does not incorporate prior knowledge of system inertia, Coriolis forces, gravity forces, friction forces and external disturbances terms. Thus, the proposed scheme does not rely on prior knowledge of these system parameters to guarantee PTC, which renders it very robust in the presence of uncertainties affecting the unknown EL system \eqref{eq:ptcuel_eom}.} {\label{rem:ctrl_policy}} 
\subsection{Stability Analysis}

This subsection demonstrates tracking errors converge to smaller bounds uniformly in prescribed-time where the states of the system \eqref{eq:ptcuel_eom} are confined to state constraints set under input constraints.

\Theorem{Consider the system \eqref{eq:ptcuel_eom} with \textit{Properties} \ref{prop:skew}--\ref{prop:matrix_bounds} and \textit{Assumptions} \ref{assume:system_matrices} {--} \ref{assume:traj_bound}. {Suppose for a sufficiently large chosen prescribed settling time $T$ as in \emph{Lemma} \ref{ptcuel_Lemma}, and $(\q, \qd) \in{\Set}$ as in \emph{Theorem} \ref{ptcuel_feasibility_theorem}, then the adaptive control policy \policy{} achieves robust local PTPB convergence to the prescribed-bound $\epsilon = \paranthesis{(\varpi/K_{p_{\Min}})^{p} + \paranthesis{\varpi\paranthesis{K_{p_\Max}/K_{p_\Min}+1}}^p}^{1/p}$ in presence of state constraints, i.e. $\bs{x}\in\mathcal{X}$ and input constraints, i.e. $\bs{u}\in\mathcal{U}$. Moreover, the adaptive gain ${K}(t)$ remains bounded for all $t\geq 0$}.}{\label{ptcuel_stability_theorem}}

\emph{Proof}: Using \eqref{eq:ptcuel_eom}, the time derivative of the \eqref{eq:ptcuel_filteredError} becomes
\begin{equation}
    \begin{aligned}
        \bs{M}\dot{\bs{\chi}} &= \bs{M}\bs{\ddot{\varepsilon}} + \bs{M} \bs{K}_p\dot{\bs{\varepsilon}} = \bs{u} - \bs{C}\bs{\chi} + \bs{\delta},\\
    \end{aligned}
    \label{eq:ptcuel_errorDynamics}
\end{equation}
where the lumped uncertain term $\bs{\delta}(t)$ is written as
\begin{equation}
\begin{aligned}[b]
    \bs{\delta}(t) &= \bs{C}(\q,\qd)(\bs{\chi} - \qd) + \bs{M}(\q)(-\dqdd - \ddot{\bs{e}}^d + \bs{K}_p\dot{\bs{\varepsilon}})\\
                   &\quad - \bs{G}(\q) - \bs{F}(\qd) - \bs{d}(t).
\end{aligned}
    \label{eq:ptcuel_lumped_uncertainy}
\end{equation}

Then, the upper bound on the lumped state uncertainty term $\bs{\delta}(t)$ can be obtained using \eqref{eq:ptcuel_filteredError}, \textit{Properties} \ref{prop:skew}--\ref{prop:matrix_bounds} and \assumeref{\ref{assume:bounded_extn}} as in \eqref{eq:ptcuel_uncertaintyBounds}.
\begin{flalign}
    \norm{\bs{\delta}} &{\leq} \norm{\bs{C}} \norm{\bs{\chi} {-} \dot{\bs{q}}} {+} \norm{\bs{M}} \norm{{-}\bs{\ddot{q}}_{d} {-} \bs{\ddot{\err}}^d {+} \bs{K}_p\dot{\bs{\varepsilon}}} {+} \norm{\bs{F}} {+}\ol{G}{+}\ol{d}\nonumber\\
    &{\leq}  \ol{C} \norm{\dot{\bs{q}}} \norm{\bs{\chi} {-} \dot{\bs{q}}} {+} \ol{M} \norm{{-}\bs{\ddot{q}}_{d} {-} \bs{\ddot{\err}}^d {+} \bs{K}_p\dot{\bs{\varepsilon}}} {+} \ol{F}\norm{\dot{\bs{q}}}{+} \ol{G}  {+} \ol{d} \nonumber\\
    &\leq \ol{\mu}\Gamma(\bs{\varepsilon}, \dot{\bs{\varepsilon}}),
    \label{eq:ptcuel_uncertaintyBounds}
\end{flalign}
 where $\ol{\mu}$ is defined as follows: 
 \begin{flalign}
    \ol{\mu} &{=} \Max\{\mu_0, \mu_1, \mu_2, \mu_3\},\ \mu_0 {=}  \ol{C} K_{P_{\Max}},  \nonumber\\
    \mu_1 &{=}\ol{M}(\ol{q}_{rdd} {+} \ol{e}_{dd}) {+} \ol{C} (\ol{q}_{rd}{+}\ol{e}_{d})^2 {+} \ol{F}( \ol{q}_{rd}{+}\ol{e}_{d}) {+} \ol{G} {+} \ol{d},\nonumber\\
    \mu_2 &{=} \ol{C}(\ol{q}_{rd} {+} \ol{e}_{d})K_{P_{\Max}},\ \mu_3 {=}  \ol{M} K_{P_{\Max}} {+} \ol{C} (\ol{q}_{rd} {+} \ol{e}_{d}) {+} \ol{F}.
\label{eq:ptcuel_parameter_bounds}
\end{flalign}
Note that the state-dependent uncertainty bound arises as a consequence of  \propref{\ref{prop:skew}}--\ref{prop:matrix_bounds} of the EL system, which is independent of the system's anatomy under consideration.

Secondly, we now prove the system is stable under the control policy in \policy{}. To this end, consider the following Lyapunov candidate:
\begin{equation}
    V(t) = \frac{1}{2}\bs{\chi}^\top\bs{M}\bs{\chi} + \frac{1}{2\gamma}\bs{\xi}^\top\bs{\xi}.
    \label{eq:ptcuel_lyapunovCandidate}
\end{equation}
Then, using the time derivative of \eqref{eq:ptcuel_lyapunovCandidate}, we have, 
\begin{equation}
    \begin{aligned}[b]
        \dot{V} &{=} \frac{1}{2}\bs{\chi}^\top\dot{\bs{M}}\bs{\chi} + \bs{\chi}^\top\bs{{M}}\dot{\bs{\chi}} + \frac{1}{\gamma}\bs{\xi}^\top\dot{\bs\xi} \\
        &{=} \bs{\chi}^\top(\bs{\Pi}\bs{\tau} {+} \bs{\delta}) {+} \frac{1}{2}\bs{\chi}^\top(\dot{\bs{M}}{-}2\bs{C})\bs{{\chi}} {+} \frac{1}{\gamma}\bs{\xi}^\top\dot{\bs{\xi}}.
    \end{aligned}
    \label{eq:ptcuel_Vdt_step1}
\end{equation}
Now, from (\ref{eq:ptcuel_Yupdate}),  
\begin{eqnarray}
\label{zeta_updatk_1}
\dot{\bs{y}}=2\bs{\Lambda}\dot{\bs{\Upsilon}}= -\gamma \bs{\xi} + \gamma\alpha \bs{D}\bs{\chi} + \dot{\bs{\Phi}} - \bs{D}\dot{\bs{\chi}}.
\end{eqnarray}
Then, by taking the time-derivative of \eqref{eq:ptcuel_StateConstrainEquation} and substituting (\ref{zeta_updatk_1}), it follows that
\begin{equation}
    \dot{\bs{\xi}} = -\gamma \bs{\xi} + \gamma\alpha \bs{D}\bs{\chi}.
    \label{eq:ptcuel_StateConstrainLaw}
\end{equation}
Invoking \propref{\ref{prop:skew}}, \policy{} and \eqref{eq:ptcuel_Vdt_step1}, we have,
\begin{flalign}
        \dot{V} {=} &{-}\frac{\varrho\Gamma(\bs{\varepsilon}, \dot{\bs{\bs{\varepsilon}}})}{\varpi{-}\norm{\bs{\chi}}}\bs{\chi}^\top\bs{\Pi} \bs{\chi} {+} \bs{\chi}^\top\bs{\delta}\nonumber \\
        & {-}  \frac{\varrho\norm{\bs{\xi}}\norm{\bs{D}}}{\varpi{-}\norm{\bs{\chi}}}\bs{\chi}^\top\bs{\Pi}\bs{\chi} {-}{\bs{\xi}}^{\top}{\bs{\xi}} {+} \alpha{\bs{\xi}}^{\top}{\bs{D}}{\bs{\chi}}.
    \label{eq:ptcuel_Vdt_step2}
\end{flalign}
According to the density property of real numbers \cite[Theorem 1.20] {rudin}, $\exists\ r > 0$, such that $0 < r \leq \lambda_{\Min}\{\bs{\Pi}\} < 1$. Besides, from \policy{}, we have $\Gamma(\bs{\varepsilon}, \dot{\bs{\bs{\varepsilon}}}) \geq 4$. Then, it follows from \eqref{eq:ptcuel_uncertaintyBounds} and \eqref{eq:ptcuel_Vdt_step2} that
\begin{equation}
    \begin{aligned}[b]
         \dot{V}{\leq} {-}{4}\paranthesis{\frac{r\varrho \norm{\bs{\chi}}}{\varpi{-}\norm{\bs{\chi}}}{-}\ol{\mu}}\!\norm{\bs{\chi}}{-}\! \paranthesis{\frac{r\varrho \norm{\bs{\chi}}}{\varpi-\norm{\bs{\chi}}}{-}\alpha}\!\norm{\bs{D}}\norm{\bs{\xi}}\norm{\bs{\chi}}.
    \end{aligned}
    \label{eq:ptcuel_Vdt_step3}
\end{equation} 
Then, provided $\norm{\bs{\chi}} \geq \Max \left \{\frac{\ol{\mu}\varpi}{r\varrho + \ol{\mu}},\ \frac{\alpha\varpi}{r\varrho + \alpha}\right\}$, we can derive that $\dot{V} \leq 0$. As $\bs{\chi}(t_0, \bs{x}(t_0)) = \bs{0}_n$, it follows that 
\begin{eqnarray}
    \label{eq:ptcuel_chi_bound}
    \norm{\bs{\chi}} \leq \Max \left \{\frac{\ol{\mu}\varpi}{r\varrho + \ol{\mu}},\ \frac{\alpha\varpi}{r\varrho + \alpha}\right\} < \varpi,\,\forall t \geq 0.
\end{eqnarray}
{Note that from \eqref{eq:ptcuel_chi_bound}, one can deduce that $\norm{\bs{\chi}}{\leq} a\varpi$ with $a{=}\Max\{\ol{\mu}/(r\varrho+\ol{\mu}), \alpha/(r\varrho+\alpha)\}{<}1$, clearly the adaptive gain $K(t)$ in \policy{} has an upper bound of $\varrho a/(1-a)$. Thus, the adaptive gains $K(t)$ in \policy{} is non-singular and bounded for all $t\geq 0$.}

Now, from \eqref{eq:ptcuel_filteredError}, the solution for the tracking error $\err$ can be found as:
\begin{flalign}
        \err &= e^{{-}\bs{K}_p{(t-t_0)}}\err_0{+}\int_{t_0}^t e^{{-}\bs{K}_p(t{-}s)}(\bs{\chi}{+}\ed^d {+} \bs{K}_p\err^d)ds \nonumber\\
             &= \int_{t_0}^t e^{{-}\bs{K}_p(t-s)}\bs{\chi} ds + \err^d.
    \label{eq:ptcuel_error_bounds_step_1}
\end{flalign}
As $\norm{\bs{\chi}} < \varpi$, $\norm{\err^d} = 0$, $\forall t\geq t_0+T$, and taking the advantage of $\bs{K}_p$ as diagonal matrix,  the bounds on tracking error are obtained $\forall t\geq t_0+T$ as follows:
\begin{flalign}
       \norm{\err} &\leq \int_{t_0}^t \norm{e^{-\bs{K}_p(t-s)}}\norm{\bs{\chi}} ds {+} \norm{\err^d} {<} \varpi\!\!\int_{t_0}^t\!\!\norm{e^{-\bs{K}_p(t-s)}} ds \nonumber\\
        &= \varpi\!\!\int_{t_0}^t\!\!e^{-K_{p_\Min}(t-s)}ds < \frac{\varpi}{K_{p_\Min}}.
    \label{eq:ptcuel_error_bounds_step_2}
\end{flalign}

Since the terms $\err^d(t)$ and $\ed^d(t)$ vanishes $\forall\,t\geq t_0+T$, it follows from \eqref{eq:ptcuel_filteredError} that $\ed(t)=\bs{\chi}-\bs{K}_p\err$. Then, the upper bound on the $\norm{\ed}$ is obtained as follows:
\begin{flalign}
    \norm{\ed(t)} {<} ||\bs{\chi}||{+}||\bs{K}_p|| ||\err||{<} \varpi\paranthesis{1 {+} K_{p_{\Max}}/K_{p_{\Min}}}.
    \label{eq:ptcuel_ultimate_rate_bound}
\end{flalign}
Thus, $\norm{[\err^\top, \ed^\top]}{\leq}\paranthesis{(\varpi/K_{p_{\Min}})^{p} {+} \paranthesis{\varpi\paranthesis{K_{p_\Max}/K_{p_\Min}{+}1}}^p}^{1/p}$ $= \epsilon$. {Thus, in line with the \emph{Definition} \ref{def:robust_ptpb_def}, the trajectories of the system \eqref{eq:ptcuel_eom} under the proposed control policy \policy{} achieve robust local PTPB convergence under the influence of external disturbances and model uncertainty $\bs{\delta}(t)$}. 

Lastly, from \eqref{eq:ptcuel_StateConstrainLaw}, the solutions of $\bs{\xi}(t)$ can be obtained as follows: 
\begin{equation}
    \bs{\xi}(t) = \gamma\alpha\int_{t_0}^te^{-\gamma(t-s)}\bs{D}\bs{\chi}dt.
    \label{eq:ptcuel_xi_solun}
\end{equation}
Then, the bounds on $\bs{\xi}$ is obtained using $\norm{\bs{\chi}}<\varpi$, as follows:
\begin{equation}
    \begin{aligned}[b]
    \norm{\bs{\xi}} &< \gamma\alpha\int_{t_0}^t\abs{e^{-\gamma(t-s)}}\norm{\bs{D}}\norm{\bs{\chi}}dt\\
                    &< \alpha\norm{\bs{D}}\varpi\paranthesis{1-e^{-\gamma t}}< \alpha\norm{\bs{D}}\varpi,\,\forall\,t\geq t_0+T.
    \end{aligned}
\end{equation}
Thus, for the choice of $c=\varpi$, we have $\norm{\bs{\xi}} < \varpi$ as $\alpha\leq 1/||\bs{D}||$. Then, from \eqref{eq:ptcuel_StateConstrainEquation}, one can obtain that $\bs{D}\bs{\chi} - \bs{\Phi} \prec  [-c\bs{1}_n^\top\ c\bs{1}_n^\top]^\top$, which implies \eqref{eq:ptcuel_DynamicStateBoundsTBGCombined}. Thus, for this \emph{apriori} choice of the safety margin $c=\varpi$, the control policy \policy{} guarantees rigorous satisfaction of the state constraints \eqref{eq:ptcuel_DynamicStateBoundsTBGCombined}.

Consequently, in the absence of the system's knowledge, the trajectories of system \eqref{eq:ptcuel_eom} under the control policy in \policy{} achieve {\emph{robust local PTPB convergence}} subjected to state constraints and input constraints. \hfill $\blacksquare$ 

\Remark{Note that the control policy is an approximation-free and can be applied to any system satisfying EL dynamics \eqref{eq:ptcuel_eom} like DC-DC power converters \cite{EL_props}, differential-drive robots \cite{funnel_mobile_el}, quadcopter \cite{quad_EL} and robotic manipulators \cite{bruno_book}.
Moreover, notice that the requirement of initial states $\bs{x}(t_0)\in{\Set}$, as stated by \emph{Theorem} \ref{ptcuel_stability_theorem} comes from the presence of input and temporal constraints for a given system \eqref{eq:ptcuel_eom} which is demonstrated in \emph{Lemma} \ref{ptcuel_Lemma} and \emph{Theorem} \ref{ptcuel_feasibility_theorem}. However, in many practical settings, model parameters are either uncertain or not known beforehand, implying system matrices' bounds are uncertain, which results in the unreliable computation of viable set ${\Set}$. In such scenarios, consistent with Remark \ref{rem:ctrl_policy}, the proposed control policy \policy{} can still be implemented by choosing a sufficiently large prescribed-time $T$ for guaranteed {robust local PTPB convergence}.}{\label{{rem:versatility}}}

\Remark {The gain matrix $\bs{K}_p$ may be chosen suitably large, and $\varpi$ suitably small, according to the inequality, $\norm{\err(t)} < \varpi/K_{p_{\Min}}$, so the tracking error $\err(t)$ may be driven to an arbitrarily small bound around the origin. However, an overly small value of $\varpi$ may result in control signal chattering. On the other hand, from (\ref{eq:ptcuel_chi_bound}), a larger value of the controller gain, $\varrho$, ensures convergence to an arbitrarily lower bound, which is, however, offset by a larger controller effort needed in achieving this tracking performance.} {\label{rem:gains}}

\Remark{{Note that from \emph{Theorem} \ref{ptcuel_stability_theorem}, the adaptive gain $K(t)$ in \policy{} has an upper bound of $\varrho a/(1-a)$. By choosing the gain $\varrho$ large enough, the value of $1-a$ increases resulting in less conservative control gains. Thus, the control gain never goes to $\infty$ and remains bounded for all $t{\geq} t_0$ while guaranteeing PTC, which is in contrast to the other studies \cite{song:2017, 
krishnaMurthy:dynamicHighGain:2020, PTC_perturbed_EL}, \cite{bertingo_7dof}}{\label{rem:tbg}}}

\Remark {Note that the adaptive barrier-function based strategy \policy{} behaves differently from the barrier function based strategy in \cite{Obeid:2018} in two key aspects: (i) it avoids introducing a discontinuity at $\boldsymbol{\chi}(t)=\bs{0}_n$, and (ii) prescribed-time convergence is achieved by relying on the TBG \eqref{eq:ptcuel_tbg_h1}, \eqref{eq:ptcuel_tbg_h2} that precludes the need for adopting a switching gain strategy. Thus, in contrast with \cite{Obeid:2018}, this policy realizes prescribed-time convergence of the error through smooth and bounded control action that alleviates chattering, while remaining bounded at the prescribed-time, $T$.} {\label{rem:adap}}

\Remark {In contrast with prevalent schemes in adaptive control literature, note that the proposed scheme \policy{} provides online adaptation and rejection of the state-dependent disturbance term, $\boldsymbol{\delta}(t)$. In particular, unlike the adaptive schemes proposed in \cite{Obeid:2020,Obeid:2018,adaptivePI:2017}, the adaptive barrier function strategy assures uniform convergence in user-prescribed finite time to a perturbation-independent ultimate bound, $\varpi/K_{p_{\Min}}$. Moreover, unlike the studies \cite{sun_Neural, Huang2023, yeCao} that achieve PTPB tracking in the presence of either state or input constraints, the proposed approximation-free control policy \policy{} achieves {robust local} PTPB convergence in the presence of both state constraints $\bs{x}\in\mathcal{X}$ and input constraints $\bs{u}\in\mathcal{U}$. } {\label{rem:novelty}}

\section{Results and Discussion} \label{sec:results}
This section presents numerical validation studies of trajectory tracking with the proposed scheme \policy{} under SIT constraints for three different robot systems. A comprehensive quantitative and qualitative comparison study is further undertaken with related studies to substantiate the superior performance of the proposed approach. In addition, numerical simulation studies are undertaken in MATLAB\textsuperscript{\textregistered} R2023a.
\subsection{Feasibility verification} \label{sec:ptcuel_feasible}

\renewcommand{\arraystretch}{1.2}
\begin{table}[!t]
    \centering
    \caption{Viable set ${\Set}$, minimum control authority $u_\Min$, and maximum disturbances $\ol{d}$, that control can handle two robots with $\qd(0)=\bs{0}_n$ and origin as the reference point in their respective joint space.}
    \begin{tabular}{|
    >{\centering\arraybackslash}m{8.5em}|
    >{\centering\arraybackslash}m{4.5em}|
    >{\centering\arraybackslash}m{4.5em}|
    >{\centering\arraybackslash}m{4.5em}|}
        \hline
        \multicolumn{4}{|c|}{R2 ($u^*=35.4\text{Nm},\ \nu^+_q=1.4810 \text{rad/s}$)} \\[1.25pt] \hline
         Variable & $T = 1.7$s & $T=2.7$s & $T=3.2$s \\ \hline
         $\norm{\q(t_0)-\q^*}$ (deg) & $30^\circ$ & $75^\circ$ & $105^\circ$\\ \hline
         $u_\Min$ (Nm) & 13.3 & 18.6 & 21.3 \\\hline
         $\ol{d}$ (Nm) & 20.0 & 15.7 & 8.5  \\\hline
         \multicolumn{4}{|c|}{IIWA 14 ($u^*=227.6\text{Nm},\ \nu^+_q=2.7706 \text{rad/s}$)}  \\[1.25pt]\hline
         Variable & $T = 2.5$s & $T=3.7$s & $T=5.0$s \\ \hline
         $\norm{\q(t_0)-\q^*}$ (deg) & $30^\circ$ & $67.5^\circ$ & $120^\circ$ \\ \hline
         $u_\Min$ (Nm) & 78.2 & 94.2 & 109.8 \\\hline
         $\ol{d}$ (Nm) & 140.8 & 108.9 & 94.7 \\\hline
    \end{tabular}
    \label{tab:ptcuel_feasible_metrics}
\end{table}
\renewcommand{\arraystretch}{1.0}
\begin{figure}[!t]
    \centering
    \includegraphics[width=0.4\textwidth]{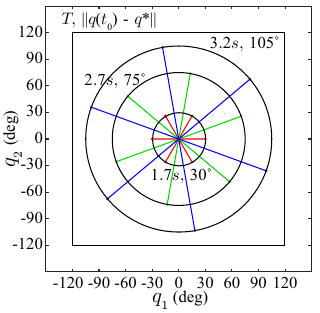}
    \caption{Simulation results show the R2 robot's performance in tracking the origin in configuration space for randomly chosen initial conditions within the viable set (refer to Table \ref{tab:ptcuel_feasible_metrics}).}
    \label{fig:ptcuel_planar_setPoint}
\end{figure}
\begin{figure}[!t]
    \centering
    \includegraphics[width=0.45\textwidth]{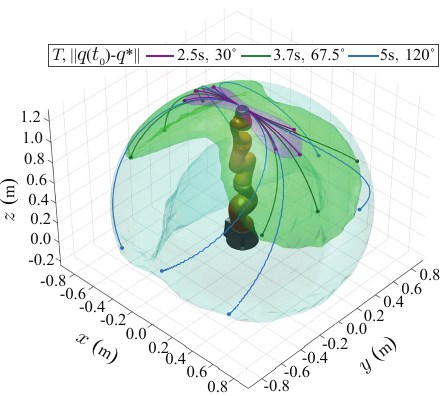}
    \caption{Simulation results show the IIWA14 robot's performance in tracking the set-point $(0,0,1.306)$ in Cartesian space for randomly chosen initial conditions within the viable set (refer to Table \ref{tab:ptcuel_feasible_metrics}). }
    \label{fig:ptcuel_iiwa14_setPoint}
\end{figure}

This subsection illustrates the choice of prescribed-time $T$ for the computation of viable set ${\Set}$ given in \eqref{eq:ptcuel_analytical_viable_set} given the state and input constraints for two different robots. We consider the origin in joint space, i.e. $\q^*=\bs{0}_n$ as reference point and set $\qd(0)=\bs{0}_n$ for the simulation study considered in this subsection. As is apparent from Table \ref{tab:ptcuel_feasible_metrics}, and consistent with Remark 1, a larger choice of the prescribed-time $T$ clearly results in a larger viable set $X_0$ for the 2-DOF planar robotic manipulator (R2) \cite{yeCao} and the 7-DOF KUKA LBR IIWA 14 R820 robot. For this simulation study, we compute the viable set ${\Set}$ and $u_\Min$ according to \eqref{eq:ptcuel_analytical_viable_set} and \eqref{eq:ptcuel_min_control_authority}. A conservative estimate of minimum control authority $u_\Min$ is obtained by using the maximum velocity of each joint pertaining to the respective feasible region, which is computed as $\norm{\q}/T$, implying the ${\nu}^+_q =\sqrt{n}\norm{\q}/T$.

Figure \ref{fig:ptcuel_planar_setPoint} illustrates the three viable regions depicted as a ball of radius $30^\circ$, $75^\circ$, and $105^\circ$ centred at the origin in the joint space of the R2 robot for prescribed-time $1.7$s, $2.7$s, and $3.2$s, respectively. For simplicity, each viable region is unshaded, but it is to be noted that the interior of the respective ball, including the boundary, represents the entire feasible region. Further, the Fig. \ref{fig:ptcuel_planar_setPoint} illustrates that the R2 robot tracks the origin from randomly selected initial points from the respective viable region within the prescribed-time, as mentioned in Table \ref{tab:ptcuel_feasible_metrics}. The innate behaviour of the traced path being linear comes from the fact that the TBG function forces the tracking error to reach the prescribed-bound set ${\mathcal{S}_\epsilon(\bs{x}_r)}$ within the user-defined settling time, which is similar to the minimum time path.

In the case of the IIWA 14 robot, the Fig. \ref{fig:ptcuel_iiwa14_setPoint} illustrates three viable regions depicted in Cartesian space corresponding to the joint space with a ball of radius $30^\circ$, $67.5^\circ$, and $120^\circ$ centred at the origin (correspondingly $P = (0,0,1.306)$m) with prescribed-time $2.5$s, $3.7$s, and $5$s respectively. For the construction of each viable region, we perform a Monte Carlo simulation by setting $\norm{\q}$ with a respective ball of radius as mentioned in Table \ref{tab:ptcuel_feasible_metrics}, followed by generating one million random points in joint space. Then, we compute the forward kinematics for the generated points to render the viable surface in Cartesian space. Fig. \ref{fig:ptcuel_iiwa14_setPoint} illustrates that the IIWA 14 robot tracks point $P$ within the prescribed-time for the randomly chosen initial points within the respective viable region. Consequently, the simulation results for both robots indicate the superior performance of the proposed scheme \policy{} that tackles the tracking performance in the presence of SIT constraints.

\subsection{Numerical Simulations} \label{sec:ptcuel_numSim}
This subsection demonstrates the efficacy of the proposed method \policy{} through numerical simulations using the KUKA LBR IIWA 14 R820 (IIWA 14) 7-DoF and the Universal Robot UR5e 6-DoF robotic manipulators to accomplish trajectory tracking within the user-defined prescribed-time. {In addition, the numerical simulations with measurement noise is also considered for two different robotic manipulators.} From the implementation standpoint, the Denavit-Hartenberg parameters of the rigid links are used to generate joint-space trajectories using an inverse kinematics algorithm \cite{rajpal_ZNN} from the Cartesian space trajectories.

\begin{table}[!t]
    \centering
    \caption{Maximum percentage steady-state error of end-effector position in Cartesian space for the robots IIWA 14 and UR5e with $T = 4\text{s}$ and $T = 3\text{s}$, respectively.}
    \begin{tabularx}{\columnwidth} {
    | >{\centering\arraybackslash}X
    | >{\centering\arraybackslash}X
    | >{\centering\arraybackslash}X
    | >{\centering\arraybackslash}X 
    |}
        \hline
        Robot & X (\%) & Y (\%) & Z (\%) \\ \hline
        IIWA 14 & 0.09 & 1.90 & 0.08 \\\hline
        UR5e & 0.04 & 0.12 & 0.14 \\\hline
    \end{tabularx}
    \label{tab:ptcuel_cartesianSpace}
\end{table}

\begin{table}[!t]
    \centering
    \caption{Maximum absolute steady-state error of joint angular position and velocity for the robots IIWA 14 and UR5e with user-defined time $T = 4\text{s}$ and $T = 3\text{s}$, respectively.}
    \begin{tabular}{|
    >{\centering\arraybackslash}m{4em}|
    >{\centering\arraybackslash}m{2em}|
    >{\centering\arraybackslash}m{2em}|
    >{\centering\arraybackslash}m{2em}|
    >{\centering\arraybackslash}m{2em}|
    >{\centering\arraybackslash}m{2em}|
    >{\centering\arraybackslash}m{2em}|
    >{\centering\arraybackslash}m{2em}|}
        \hline
        \multicolumn{8}{|c|}{Joint Angular Position Error $\times 10^{-3}$ (degree) } \\\hline
         Robot & $q_1$  & $q_2$  & $q_3$  & $q_4$  & $q_5$  & $q_6$  & $q_7$  \\ \hline
         IIWA 14  & 1 & 39 & 10 & 47 & 4 & 4.9 & 0.1 \\\hline
        UR5e  & 3 & 40 & 28 & 10 & 2 & 3& -
 \\\hline
        \multicolumn{8}{|c|}{Joint Angular Velocity Error $\times 10^{-1}$ (degrees/s) } \\\hline
        Robot & $\dot{q}_1$  & $\dot{q}_2$  & $\dot{q}_3$  & $\dot{q}_4$  & $\dot{q}_5$  & $\dot{q}_6$ &  $\dot{q}_7$ \\ \hline
         IIWA 14 & 3  & 2.4  & 4.3  & 8.7  & 3.8  & 27.5  & 5 \\\hline
        UR5e & 1 & 4.3 & 10.6 & 23.0 & 7.1 & 49.6 & -  \\\hline
    \end{tabular}
    \label{tab:ptcuel_trackingError}
\end{table}

To begin, the numerical simulation study considers the tricuspid trajectory for a duration of $30\text{s}$. An offset of $30^\circ$ from the desired joint angular position and zero joint angular velocity at $t = 0$ are considered as initial states of the system. The user-defined constants for the IIWA 14 robot are $\bs{K}_p = \text{diag}\paranthesis{[1600,\ 8000,\ 2200,\ 4000,\ 800,\ 1200,\ 128]},\ \varrho = 10.5,\, \varpi = 25.0,\ \gamma = 1,\ \alpha = 0.4,\ T = 4\text{s}$ and the state and input constraints are $\bs{\theta}_q^+ = -\bs{\theta}_q^- = (2\pi/3)\bs{1}_7\ \text{rad},\ \bs{\nu}_q^+ = - \bs{\nu}_q^- =  (\pi/6)\!\cdot\!\bs{1}_7\ \text{rad/s}\ \text{and}\ \bs{\ol{u}} = [100\!\cdot\!\bs{1}_5^\top,\ 30\!\cdot\!\bs{1}_2^\top]^\top\ \text{Nm}$, respectively. On the other hand, the user-defined constants for UR5e robot are $\bs{K}_p = \text{diag}\paranthesis{[400,\ 3600,\ 2400,\ 400,\ 400,\ 1200]},\ \varrho = 4.5,\, \varpi = 5.0,\ \gamma = 1,\ \alpha = 0.4 ,\ T = 3\text{s}$, and the state and input constraints are $\bs{\theta}_q^+ = -\bs{\theta}_q^- = (2\pi/3)\!\cdot\!\bs{1}_6\ \text{rad},\ \bs{\nu}_q^+ = - \bs{\nu}_q^- = (\pi/6)\!\cdot\!\bs{1}_6\ \text{rad/s} \ \text{and}\ \bs{\ol{u}} = \text{diag}\paranthesis{[100,\ 100,\ 100,\ 25\ 25,\ 0.1]}\ \text{Nm}$, respectively. In addition, the simulation study considers the external disturbance drawn from uniform distribution with the maximum value of $\text{abs}\paranthesis{\bs{d}(t)} = [0.01,\ 0.2,\ 0.2,\ 0.2,\  0.01,\ 0.01,\  0.001]^\top\ \text{Nm}$ for the IIWA 14 robot and $\text{abs}\paranthesis{\bs{d}(t)} = [0.01,\ 0.2,\ 0.2,\ 0.2,\  0.01,\  0.001]^\top\ \text{Nm}$ for the UR5e robot.

\begin{figure}[!t]
    \centering
    \includegraphics[width=0.45\textwidth]{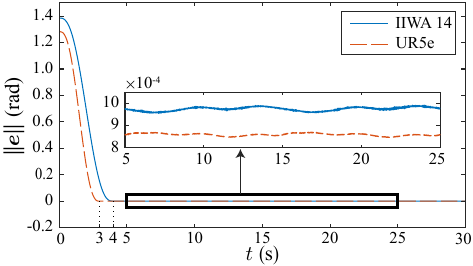}
    \caption{Simulation results for the norm of trajectory tracking error $\norm{\err}$, for the IIWA 14 and the UR5e robots with prescribed-time $T = 4\text{s and}\ T = 3\text{s} $, respectively.}
    \label{fig:ptcuel_r_norm}
\end{figure}

Table \ref{tab:ptcuel_cartesianSpace} depicts the steady-state error metrics for both robots using the proposed scheme. Clearly, the tracking error in Cartesian space is significantly low, indicating that the corresponding joint space trajectory tracking performance is highly precise, which can be seen from Table \ref{tab:ptcuel_trackingError}. The error metrics in Table \ref{tab:ptcuel_trackingError} shows the proposed controller \policy{} can drive the tracking errors within the prescribed-time to very small values that lie well within the prescribed-bound of $\varpi/K_{p_{\Min}} = 11.2^\circ$ for the IIWA 14 and  $\varpi/K_{p_{\Min}} = 0.7^\circ$ for the UR5e. Note that the ultimate bound $\varpi/K_{p_\Min}$ for IIWA14 is relatively high due to the gains we have chosen, but the tracking error $\norm{\err}$, for the IIWA14 is in the order of $0.01^\circ$ as seen from Fig. \ref{fig:ptcuel_r_norm}. Further, it is worth noticing that the joint angular position errors are not greater than $0.05^\circ$ for each robot. In addition, Fig. \ref{fig:ptcuel_r_norm} depicts that even for a large initial offset from the desired trajectory, tracking error converges to very small values in the prescribed-time $T = 4\text{s}$ and $T = 3\text{s}$ for the IIWA 14 and UR5e robots,  respectively. On the other hand, the joint angular velocity errors are less than $3^\circ /\text{s}$ for the IIWA 14 robot and $5^\circ/\text{s}$ for the UR5e robot, which is consistent with the tracking error criterion $\norm{\ed} < \varpi(1 + K_{p_{\Max}}/K_{p_{\Min}})$ obtained in \eqref{eq:ptcuel_ultimate_rate_bound}.

\begin{figure*}[!t]
    \centering
    \subfloat[]{\includegraphics[width=0.325\textwidth]{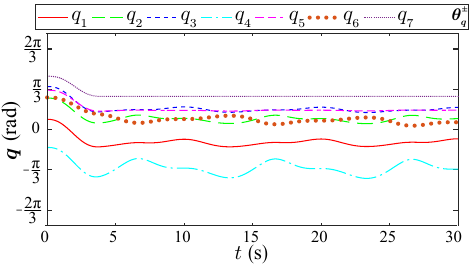}%
    \label{fig:ptcuel_iiwa14_jointPosition}}
    \hfil
    \subfloat[]{\includegraphics[width=0.325\textwidth]{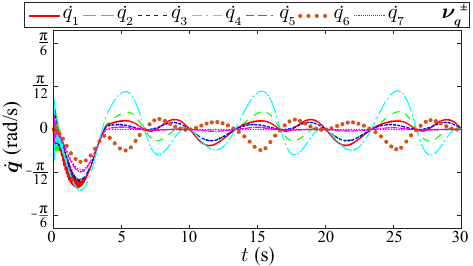}
    \label{fig:ptcuel_iiwa14_jointVelocity}}
    \hfil
    \subfloat[]{\includegraphics[width=0.325\textwidth]{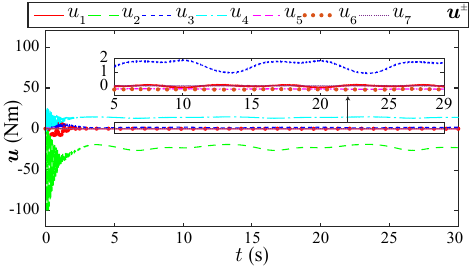}
    \label{fig:ptcuel_iiwa14_jointTorque}}
    \caption{Simulations results for the IIWA 14 robot for trajectory tracking of a tricuspid path with the proposed control policy \policy{} and prescribed-time $T = 4\text{s}$. (a) joint angular position $(\q)$, (b) joint angular velocity $(\qd)$,  and (c) control input $(\bs{u})$.}
\end{figure*}

\begin{figure*}[!t]
    \centering
    \subfloat[]{\includegraphics[width=0.325\textwidth]{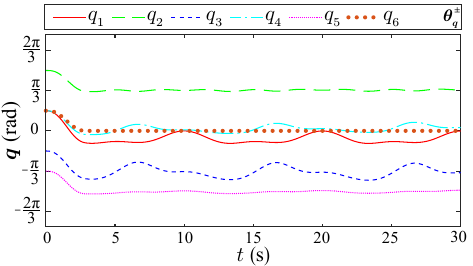}%
    \label{fig:ptcuel_ur5e_jointPosition}}
    \hfil
    \subfloat[]{\includegraphics[width=0.325\textwidth]{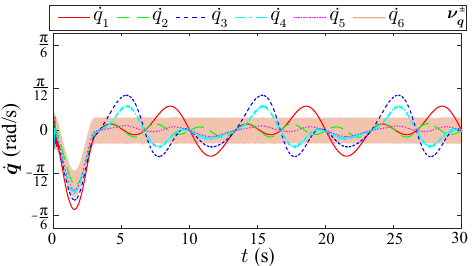}
    \label{fig:ptcuel_ur5e_jointVelocity}}
    \hfil
    \subfloat[]{\includegraphics[width=0.325\textwidth]{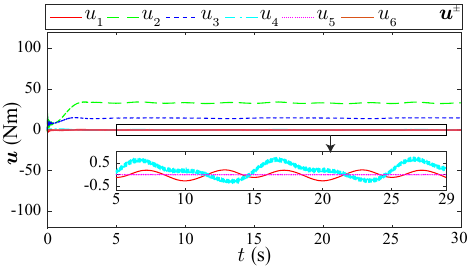}
    \label{fig:ptcuel_ur5e_jointTorque}}
    \caption{Simulations results for the UR5e robot for trajectory tracking of a tricuspid path with the proposed control policy \policy{} and prescribed-time $T = 3\text{s}$. (a) joint angular position $(\q)$, (b) joint angular velocity $(\qd)$,  and (c) control input $(\bs{u})$.}
\end{figure*}

It is clear from Fig. \ref{fig:ptcuel_iiwa14_jointPosition} and \ref{fig:ptcuel_iiwa14_jointVelocity} that the control policy \policy{} ensures the state constraint satisfaction for the IIWA 14 robot. Similar conclusions can be drawn for the UR5e robot based on Fig. \ref{fig:ptcuel_ur5e_jointPosition} and \ref{fig:ptcuel_ur5e_jointVelocity}. From Fig. \ref{fig:ptcuel_iiwa14_jointTorque} and \ref{fig:ptcuel_ur5e_jointTorque}, it is observed that reasonably low levels of control effort are expended to ensure that tracking errors converge to the prescribed-bound $\varpi/K_{p_{\Min}}$ within the prescribed-time, $T$. In addition, the constraint transformation in \eqref{eq:ptcuel_SaturationFunction} ensures that the input constraints are satisfied without loss of the tracking performance. 


\subsubsection{With Measurement Noise}
\begin{figure*}[!t]
    \centering
    \includegraphics[width=\textwidth]{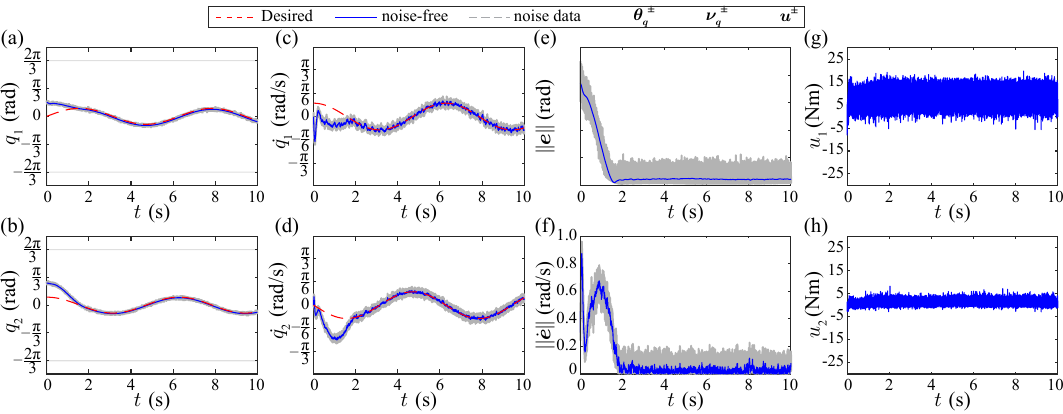}
    \caption{Simulation results for the R2 robot tracking the sinusoidal trajectory with a prescribed-time $T = 2 \text{s}$, for the proposed control scheme \policy{} with measurement noise of 30 dB. (a,b) joint angular position tracking, (c,d) joint angular velocity tracking, (e,f) norm of tracking errors $\err\ \text{and}\ \ed$, and (g,h) control input, respectively.}
    \label{fig:ptcuel_planar_jointVals_noise}
\end{figure*}

\begin{figure*}[!t]
    \centering
    \includegraphics[width=\textwidth]{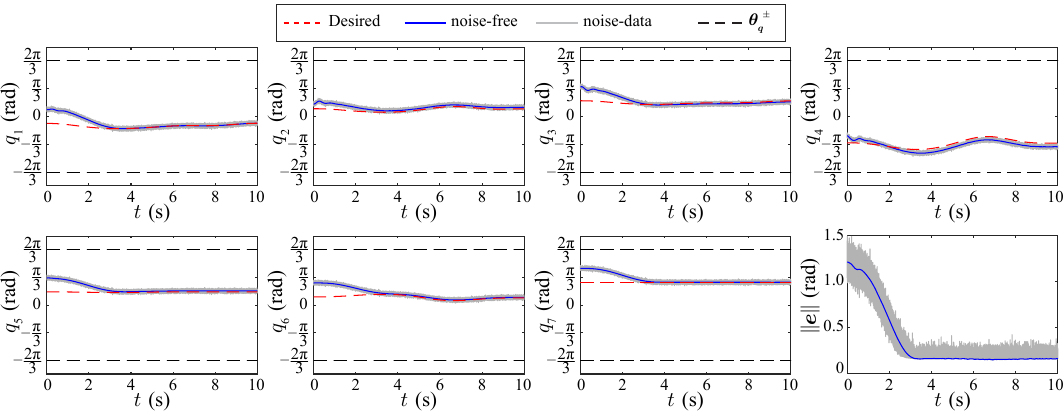}
    \caption{Simulation results for the IIWA 14 robot manipulator shows the joint angular position tracking with the prescribed-time $T = 4\text{s}$ for the proposed control scheme \policy{} with Measurement Noise 30 dB.}
    \label{fig:ptcuel_iiwa14_jointPosition_noise}
\end{figure*}

\begin{figure*}[!t]
    \centering
    \includegraphics[width=\textwidth]{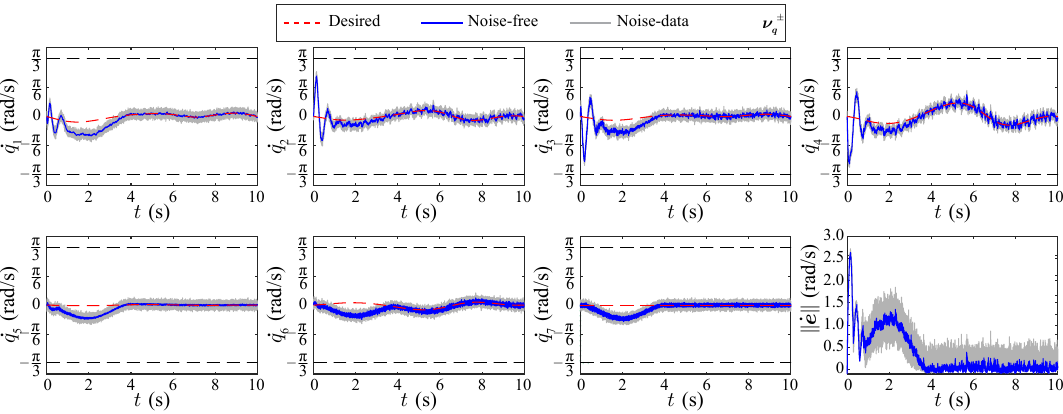}
    \caption{Simulation results for the IIWA 14 robot manipulator shows the joint angular velocity tracking with the prescribed-time $T = 4\text{s}$ for the proposed control scheme \policy{} with Measurement Noise 30 dB.}
    \label{fig:ptcuel_iiwa14_jointVelocity_noise}
\end{figure*}

{In this sub-section, we demonstrate that the proposed adaptive policy \policy{} is robust to measurement noise. In this regard, we consider sensor noise of magnitude 30 dB for each joint for two robots, namely the planar robotic manipulator (R2) with two revolute joints and the IIWA14. The controller gains are chosen as $\bs{K}_p = \text{diag}\paranthesis{[60,18]},\ \varrho = 8$ for the R2 robot and $\bs{K}_p = \text{diag}\paranthesis{[12, 80, 12, 20, 1, 2,1]}, \varrho=12.5$ for the IIWA14 robot. These gains are chosen to be lower than the gains for the noise-free case because the proposed policy \policy{} is constructed based on filtered-tracking error, which amplifies the sensor noise through the gains $\bs{K}_p$. From the error metrics in Tables \ref{tab:ptcuel_noise_error_metrics_planar_R2}-\ref{tab:ptcuel_noise_error_metrics_IIWA14_velocity},  and Figs. \ref{fig:ptcuel_planar_jointVals_noise}-\ref{fig:ptcuel_iiwa14_jointVelocity_noise}, it is apparent that even in the presence of measurement noise and external disturbances, the proposed controller \policy{} achieves robust local PTPB convergence consistent with \emph{Definition}, \ref{def:robust_ptpb_def} and \emph{Theorem} \ref{ptcuel_stability_theorem}.}

\begin{table}[!t]
    \centering
    \caption{Steady-state error metrics in joint angular position (degree) and joint angular velocity (degrees/s)  with measurement noise of 30 dB for the proposed method \policy{} computed after the prescribed-time $T = 2\text{s}$ for the R2 robot.}
    \begin{tabular}{|
    >{\centering\arraybackslash}m{6em}|
    >{\centering\arraybackslash}m{4em}|
    >{\centering\arraybackslash}m{4em}|
    >{\centering\arraybackslash}m{4em}|
    >{\centering\arraybackslash}m{4em}|}
        \hline
          Method & $q_1$ (deg) & $q_2$ (deg) & $\dot{q}_1$ (deg/s)  & $\dot{q}_2$ (deg/s)\\\hline
        \multicolumn{5}{|c|}{Maximum Absolute Error}\\\hline
          Noise-free & 2.57 & 1.46 & 5.84 & 3.92\\\hline
           Noise & 9.70 & 8.67 & 10.52 & 8.55  \\\hline
           \multicolumn{5}{|c|}{Mean Absolute Error}\\\hline
           Noise-free & 2.33 & 1.14 & 1.27 & 0.70 \\\hline
          Noise &2.50 & 1.73 & 1.92 & 1.61 \\\hline
           \multicolumn{5}{|c|}{RMSE}\\\hline
           Noise-free & 2.08 & 1.03 & 1.42 & 0.80 \\\hline
          Noise & 2.64 & 1.92 & 2.15 & 1.80 \\\hline
    \end{tabular}
    \label{tab:ptcuel_noise_error_metrics_planar_R2}
\end{table}

\begin{table}[!t]
    \centering
    \caption{Steady-state error metrics of joint angular position with measurement noise of 30 dB for the proposed method \policy{} computed after the prescribed-time $T = 4\text{s}$ for the IIWA 14 robot.}
    \begin{tabular}{|
    >{\centering\arraybackslash}m{4em}|
    >{\centering\arraybackslash}m{2em}|
    >{\centering\arraybackslash}m{2em}|
    >{\centering\arraybackslash}m{2em}|
    >{\centering\arraybackslash}m{2em}|
    >{\centering\arraybackslash}m{2em}|
    >{\centering\arraybackslash}m{2em}|
    >{\centering\arraybackslash}m{2em}|}
        \hline
          Method & $q_1$  & $q_2$  & $q_3$ & $q_4$ & $q_5$ & $q_6$  & $q_7$ \\\hline
        \multicolumn{8}{|c|}{Maximum Absolute Error (deg)} \\\hline
          Noise-Free & 0.29 & 3.75 & 2.22 & 8.31 & 1.99 & 2.16 & 0.06 \\\hline
          Noise  & 8.23 & 10.87 & 10.00 & 15.62 & 9.29 & 8.56 & 8.02 \\\hline
          \multicolumn{8}{|c|}{Mean Absolute Error (deg) }\\\hline
          Noise-Free & 0.16 & 3.34 & 1.90 & 7.62 & 1.88 & 1.30 & 0.01 \\\hline
          Noise  & 1.45 & 3.40 & 2.18 & 7.61 & 2.17 & 1.84 & 1.45 \\\hline
          \multicolumn{8}{|c|}{RMSE (deg)}\\\hline
          Noise-Free & 0.13 & 2.59 & 1.48 & 5.91 & 1.45 & 1.06 & 0.01 \\\hline
          Noise  & 1.41 & 2.95 & 2.04 & 6.07 & 2.03 & 1.76 & 1.41 \\\hline
    \end{tabular}    
    \label{tab:ptcuel_noise_error_metrics_IIWA14_position}
\end{table}

\begin{table}[!t]
    \centering
    \caption{Steady-state error metrics of joint angular velocity with measurement noise of 30 dB for the proposed method \policy{} computed after the prescribed-time $T = 4\text{s}$ for the IIWA 14 robot.}
    \begin{tabular}{|
    >{\centering\arraybackslash}m{4em}|
    >{\centering\arraybackslash}m{2em}|
    >{\centering\arraybackslash}m{2em}|
    >{\centering\arraybackslash}m{2em}|
    >{\centering\arraybackslash}m{2em}|
    >{\centering\arraybackslash}m{2em}|
    >{\centering\arraybackslash}m{2em}|
    >{\centering\arraybackslash}m{2em}|}
        \hline
          Method & $q_1$  & $q_2$  & $q_3$ & $q_4$ & $q_5$ & $q_6$  & $q_7$ \\\hline
        \multicolumn{8}{|c|}{Maximum Absolute Error (deg/s)} \\\hline
          Noise-Free & 2.07 & 5.90 & 5.30 & 9.96 & 1.38 & 5.15 & 3.15 \\\hline
          Noise  & 8.02 & 11.12 &  10.70 &  14.43 &  7.86 & 10.65 & 8.03 \\\hline
          \multicolumn{8}{|c|}{Mean Absolute Error (deg/s) }\\\hline
          Noise-Free & 0.43 & 1.27 & 0.95 & 2.03 & 0.29 & 1.12 & 0.61 \\\hline
          Noise  &1.51 & 1.92 & 1.73 & 2.49 & 1.48 & 1.84 & 1.57 \\\hline
          \multicolumn{8}{|c|}{RMSE (deg/s) }\\\hline
          Noise-Free & 0.42 & 1.23 & 0.94 & 1.98 & 0.28 & 1.10 & 0.59
          \\\hline
          Noise  & 1.47 & 1.87 & 1.69 & 2.42 & 1.43 & 1.78 & 1.52 \\\hline
    \end{tabular}    
    \label{tab:ptcuel_noise_error_metrics_IIWA14_velocity}
\end{table}

\begin{figure*}[!t]
    \centering
    \includegraphics[width=\textwidth]{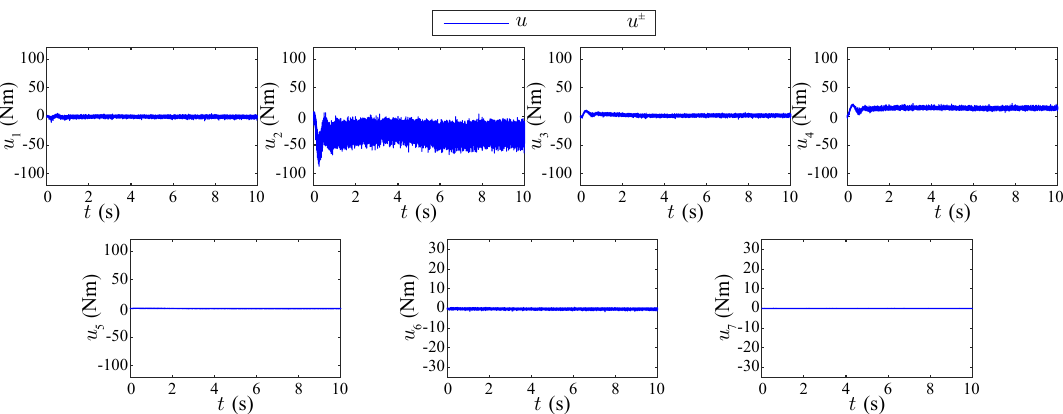}
    \caption{Simulation results for the IIWA 14 robot manipulator shows the Control Input variation with the prescribed-time $T = 4\text{s}$ for the proposed control scheme \policy{} with Measurement Noise 30 dB.}
    \label{fig:ptcuel_iiwa14_jointTorque_noise}
\end{figure*} 

\subsection{Experimental Studies}
\begin{table}[!t]
    \centering
    \caption{Experimental study of steady-state error metrics of joint angular position and velocity using the proposed policy with/without payload of 1Kg for the Franka Research 3 robot.}
    \begin{tabular}{|
    >{\centering\arraybackslash}m{3em}|
    >{\centering\arraybackslash}m{2.25em}|
    >{\centering\arraybackslash}m{2.25em}|
    >{\centering\arraybackslash}m{2.25em}|
    >{\centering\arraybackslash}m{2.25em}|
    >{\centering\arraybackslash}m{2.25em}|
    >{\centering\arraybackslash}m{2.25em}|
    >{\centering\arraybackslash}m{2.25em}|}
        \hline
          Payload & $q_1$  & $q_2$  & $q_3$ & $q_4$ & $q_5$ & $q_6$  & $q_7$ \\\hline
        \multicolumn{8}{|c|}{Maximum Absolute Steady State Error (deg) }\\\hline
          0 Kg & 0.30& 0.20& 0.24& 0.27& 0.21& 0.21& 0.16 \\\hline
          1 Kg & 0.31& 0.58& 0.26& 0.45& 0.25& 0.20& 0.18 \\\hline
          \multicolumn{8}{|c|}{Root Mean Square Error (RMSE) (deg)  }\\\hline
           0 Kg & 0.09 & 0.06& 0.10 & 0.08& 0.10 & 0.10& 0.10 \\\hline
        1 Kg & 0.08& 0.14& 0.10& 0.11& 0.13& 0.09& 0.10 \\\hline
          Method & $\dot{q}_1 $  & $\dot{q}_2 $  & $\dot{q}_3 $  & $\dot{q}_4 $  & $\dot{q}_5 $  & $\dot{q}_6 $  & $\dot{q}_7 $ \\\hline
        \multicolumn{8}{|c|}{Maximum Absolute Steady State Error (in deg/s)}\\\hline
         0 Kg &14.42& 8.52& 8.19& 11.46& 10.96& 7.65& 3.95 \\\hline
         1 Kg & 15.56& 7.74& 11.13& 12.34& 13.22& 5.66& 17.27 \\\hline
            \multicolumn{8}{|c|}{Root Mean Square Error (RMSE) (in deg/s)}\\\hline
          0 Kg & 4.32& 2.19& 2.25& 2.90& 3.77& 1.69& 0.99\\\hline
        1 Kg & 3.88& 1.93& 2.16& 3.00& 4.05& 1.58& 1.56 \\\hline
    \end{tabular}    
    \label{tab:ptcuel_frankaPosErr}
\end{table}
\begin{figure*}[!t] 
    \centering
    \subfloat[Joint Angular Position $(\q)$]{\includegraphics[width=0.35\textwidth]{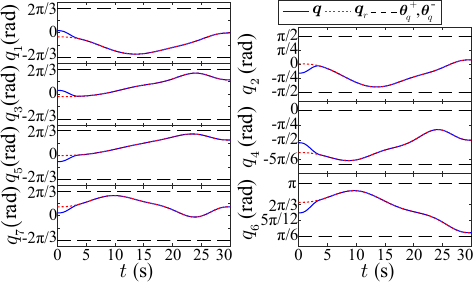}%
    \label{fig:ptcuel_franka_jointPosition}}
    \hfill
    \subfloat[Joint Angular Velocity $(\qd)$]{\includegraphics[width=0.35\textwidth]{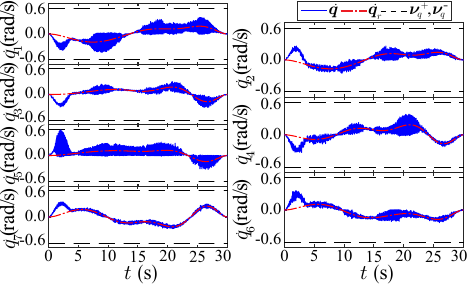}
    \label{fig:ptcuel_franka_jointVelocity}}
    \hfill
    \subfloat[Control Input $(\bs{u})$]{\includegraphics[width=0.35\textwidth]{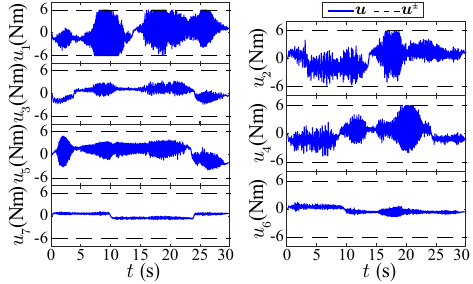}
    \label{fig:ptcuel_franka_jointTorque}}
    \label{fig:ptcuel_franka_full}
    \caption{Experimental results with the Franka Research 3 robotic arm tracking a reference trajectory illustrating the satisfaction of state and input constraints using the proposed control policy \policy{}.}
\end{figure*}
\begin{figure}
    \centering
    \includegraphics[width=0.9\linewidth]{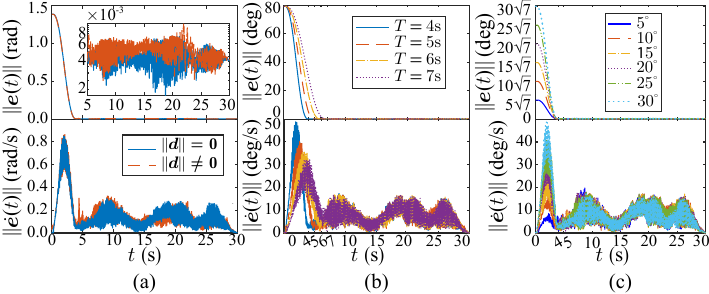}
    \caption{Experimental results with the Franka Research 3 robot shows the tracking performance for various initial conditions and prescribed settling time.}
    \label{fig:ptcuel_franka_variation}
\end{figure}

This subsection demonstrates the efficacy of the control policy \policy{} through the experimental studies. For this study, we considered the Franka Research 3 Robotic manipulator, a 7-DoF serial robotic arm. The following are the controller parameters chosen for this study: $\bs{K}_p = \text{diag}\paranthesis{[340, 324, 128, 224, 196, 64, 64]},\ \varrho = 10.5,\, \varpi = 25.0,\ \gamma = 1,\ \alpha = 0.4,\ T = 4\text{s}$ and the state and input constraints are $\bs{\theta}_q^+ = [2.094, 1.5708, 2.094, 0.0, 2.094, 3.14, 2.094]^\top\text{rad}$, $\bs{\theta}_q^- = [-2.094, -1.5708, -2.094, -2.86, -2.094, 0.5236, -2.094]^\top,\ \bs{\nu}_q^+ = - \bs{\nu}_q^- = 0.6\!\cdot\!\bs{1}_7\ \text{rad/s}\ \text{and}\ \bs{{u}}^+ = -\bs{u}^- =  0.6\!\cdot\!\bs{1}_7\ \text{Nm}$, respectively.

Firstly, an offset of $30^\circ$ for each joint angular position from the desired joint angular position and zero joint angular velocity at $t = 0$ are considered as initial states of the system. The steady-state metrics in Table \ref{tab:ptcuel_frankaPosErr} show that the maximum absolute steady-state error (MASE) and root mean square error (RMSE) for the robot are in the order of $0.1^\circ$ for joint angular position, which illustrates that the proposed control policy achieves local PTPB convergence. Moreover, a 1Kg payload is attached to the robot arm and for the same initial conditions and the reference trajectory, the robot robustly rejects the disturbances and achieves robust local PTPB convergence under the control policy \policy{} as in \emph{Definition}  \ref{def:robust_ptpb_def} as shown in Fig. \ref{fig:ptcuel_franka_variation}a and the Table \ref{tab:ptcuel_frankaPosErr}. Further, as displayed in Fig. \ref{fig:ptcuel_franka_jointPosition}, \ref{fig:ptcuel_franka_jointVelocity}, and \ref{fig:ptcuel_franka_jointTorque},  the proposed control policy also satisfies SIT constraints with $T = 4$s. Furthermore, a variation of prescribed-settling time $T = 4,5,6,7$s and a variation of initial conditions with offsets $\abs{q_i - q_{i,r}} = 5^\circ,  10^\circ,15^\circ,20^\circ,25^\circ, 30^\circ $ are demonstrated in Fig. \ref{fig:ptcuel_franka_variation}b and \ref{fig:ptcuel_franka_variation}c to illustrate that the local PTPB convergence is achieved for any user-defined settling time and any initial conditions. 

\subsection{Comparison With Related Studies}
This subsection considers a quantitative comparison of the proposed controller \policy{} with two leading alternative designs, \cite{yeCao} and \cite{Huang2023}.

A set of simulation studies is undertaken to provide a quantitative comparison with the proposed control policy \policy{} against \cite{Huang2023} and \cite{yeCao} using two robots, namely the planar robotic manipulator (R2) with two revolute joints and the KUKA LBR IIWA 14 R820 (IIWA 14) robotic manipulator. For the R2 robot, state constraints are given by $\bs{\theta}_q^+ = -\bs{\theta}_q^- = [2\pi/3,\ 2\pi/3]^\top\ \text{rad}$, $\bs{\nu}_q^+ = -\bs{\nu}_q^- = [\pi/3,\ \pi/3]^\top\ \text{rad/s}$ and input constraints are given by $\ol{\bs{u}} = [25,\ 25]^\top\ \text{(Nm)}$. The physical properties of the R2 and external disturbances formulation are extracted from \cite{yeCao}. On the other hand, for the IIWA 14, state and input constraints are given by $\bs{\theta}_q^+ = -\bs{\theta}_q^- = (2\pi/3)\!\cdot\!\bs{1}_7^\top\ \text{rad},\ \bs{\nu}_q^+ = - \bs{\nu}_q^- =  (\pi/3)\!\cdot\!\bs{1}_7^\top\ \text{rad/s}$ and $ \bs{\ol{u}} = [100\!\cdot\!\bs{1}_5^\top,\ 30\!\cdot\!\bs{1}_2^\top]^\top\ \text{Nm}$ respectively, and the external disturbances are drawn from a uniform distribution with the maximum value of $\text{abs}\paranthesis{\bs{d}(t)} = [0.01,\ 0.2,\ 0.2,\ 0.2,\  0.01,\ 0.01,\  0.001]^\top\ \text{Nm}$. Further, the desired trajectory profile is set as $\dq(t) = [0.3\sin(t),\ 0.3\cos(t)]^\top$ for the R2 robot, while desired joint space trajectories are recovered from the tricuspid trajectory in Cartesian space for the IIWA 14 as mentioned in Section \ref{sec:ptcuel_numSim} for the duration of 10 seconds. The prescribed-time parameter is set as $T = 2\text{s}$ for the R2 robot and $T = 4\text{s}$ for the IIWA 14 robot, respectively. The simulation study imposes the same initial conditions on the system states with $30^\circ$ offset from reference trajectory for joint angular position and zero joint angular velocity at $t=0$ for all the control schemes. In addition, the same limits on states and inputs are imposed for all the approaches, and these are imposed as hard limits wherever these limits are not considered in controller design, as described in Table \ref{tab:ptcuel_qualitative_comparison}. Furthermore, the gains for the proposed scheme \policy{} are chosen as $\bs{K}_p = \text{diag}\paranthesis{[1600,\ 8000,\ 2200,\ 4000,\ 800,\ 1200,\ 128]},\ \varrho = 10.5,\, \varpi = 25.0,\ \gamma = 1,\ \alpha = 0.4$, for the IIWA 14 robot, and $\bs{K}_p = \text{diag}\paranthesis{[2400,\ 1000]},\ \varrho = 8,\, \varpi = 2.0,\ \gamma = 1,\ \alpha = 0.4$. For the control scheme in \cite{Huang2023}, the gains are chosen as $\kappa_1=2, \Gamma=1, \eta=0.8, \delta=1, c=1, \varepsilon=1, \ol{\delta} = \udl{\delta} = 1, k_1=30, k_2=50, k_3=0.001, \ol{M}=0.001\bs{I}_7$ for the IIWA 14, and $\kappa_1=4, \Gamma=1, \eta=0.8, \delta=1, c=1, \varepsilon=0.1, \ol{\delta} = \udl{\delta} = 1, k_1=2, k_2=6, k_3=0.01, \ol{M}=0.005\bs{I}_2$, for the R2. For the control scheme in \cite{yeCao}, the gains are chosen as $k_1 = 32,\ k2 = 20,\ r_1 = 16,\ r_2 = 10,\ r_3 = 4,\ r_4 = 25,\ w_1 = 0.01\ \varepsilon = 2.0$, for the IIWA 14 and $k_1 = 32,\ k_    2 = 15,\ r_1 = 8,\ r_2 = 10,\ r_3 = 2,\ r_4 = 25,\ w_1 = 0.001\ \varepsilon = 2.0$, for the R2.

\begin{table}[!t]
    \centering
    \caption{Comparison of steady-state error metrics in joint angular position (degree) and joint angular velocity (degrees/s) between the proposed method \policy{} and \cite{Huang2023} and \cite{yeCao} computed after the prescribed-time $T = 2\text{s}$ for the R2 robot.}
    \begin{tabular}{|
    >{\centering\arraybackslash}m{6em}|
    >{\centering\arraybackslash}m{4em}|
    >{\centering\arraybackslash}m{4em}|
    >{\centering\arraybackslash}m{4em}|
    >{\centering\arraybackslash}m{4em}|}
        \hline
          Method & $q_1$ (deg) & $q_2$ (deg) & $\dot{q}_1$ (deg/s)  & $\dot{q}_2$ (deg/s)\\\hline
        \multicolumn{5}{|c|}{Maximum Absolute Error $\paranthesis{\times 10^{-3}}$ }\\\hline
          Proposed \policy{} & \textbf{9} & \textbf{4} & \textbf{2} & \textbf{1}\\\hline
           \cite{Huang2023} & 549 & 520 & 124 & 150  \\\hline
           \cite{yeCao}  & 1356 & 575 & 632 & 446 \\\hline
           \multicolumn{5}{|c|}{RMSE $\paranthesis{\times 10^{-3}}$ }\\\hline
           Proposed \policy{} & \textbf{8} & \textbf{3} & \textbf{0.4} & \textbf{0.3} \\\hline
          \cite{Huang2023} & 83 & 75 & 187 & 229 \\\hline
          \cite{yeCao} & 728 & 272 & 276 & 281\\\hline
    \end{tabular}
    \label{tab:ptcuel_R2}
\end{table}

\begin{figure*}[!t]
    \centering
    \includegraphics[width=\textwidth]{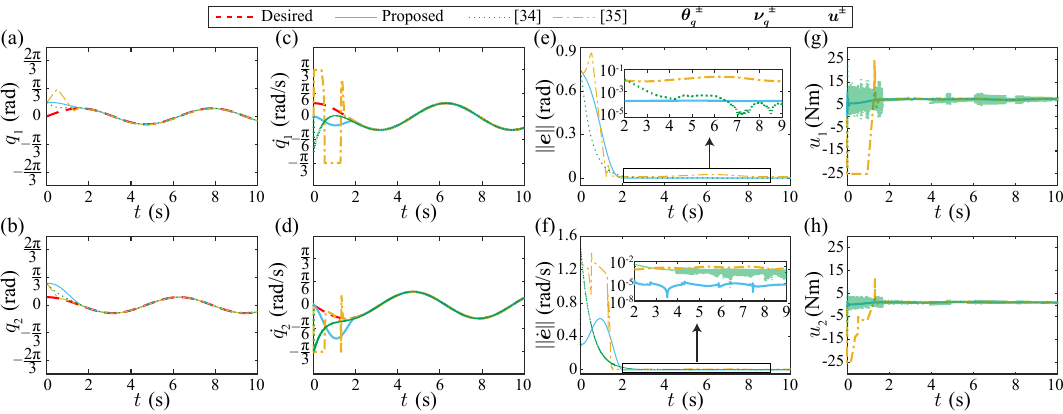}
    \caption{Simulation results for the R2 robot tracking the sinusoidal trajectory with a prescribed-time $T = 2 \text{s}$, for the proposed control scheme \policy{} against \cite{Huang2023} and \cite{yeCao}. (a,b) joint angular position tracking, (c,d) joint angular velocity tracking, (e,f) norm of tracking errors $\err\ \text{and}\ \ed$, and (g,h) control input, respectively.}
    \label{fig:ptcuel_planar_jointVals}
\end{figure*}

For R2, the error metrics in Table \ref{tab:ptcuel_R2} show that the control policy in \policy{} outperforms the two other methods \cite{Huang2023} and \cite{yeCao}.  Note that both the maximum absolute error and the root mean square error (RMSE) of the joint angular position are in the order of $10^{-3}$ degrees for the proposed controller, and $10^{-1}$ degrees using the other two methods \cite{Huang2023} and \cite{yeCao}. Also, a similar trend in the case of joint angular velocity for these methods is observed. Further, Fig. \ref{fig:ptcuel_planar_jointVals}a and \ref{fig:ptcuel_planar_jointVals}d shows the smooth transient performance (for $0 < t \leq T$) for all methods except for \cite{yeCao}, and the proposed controller has better performance than \cite{Huang2023} and \cite{yeCao} during the steady state for joint angular position tracking, as seen in Fig. \ref{fig:ptcuel_planar_jointVals}c. In the case of joint angular velocity tracking from Fig. \ref{fig:ptcuel_planar_jointVals}b and \ref{fig:ptcuel_planar_jointVals}e, it can be observed that $\qd$ saturates during the transient stage for the control policies in  \cite{Huang2023} and \cite{yeCao}, respectively. Also, the method in \cite{Huang2023} shows chattering during the steady state, as seen in Fig. \ref{fig:ptcuel_planar_jointVals}f, as this approach implements a discontinuous control policy. 

\begin{table}[!t]
    \centering
    \caption{Comparison of steady-state error metrics of joint angular position in degrees between the proposed method \policy{}, \cite{Huang2023} and \cite{yeCao} computed after the prescribed-time $T = 4\text{s}$ for the IIWA 14 robot.}
    \begin{tabular}{|
    >{\centering\arraybackslash}m{4em}|
    >{\centering\arraybackslash}m{2em}|
    >{\centering\arraybackslash}m{2em}|
    >{\centering\arraybackslash}m{2em}|
    >{\centering\arraybackslash}m{2em}|
    >{\centering\arraybackslash}m{2em}|
    >{\centering\arraybackslash}m{2em}|
    >{\centering\arraybackslash}m{2em}|}
        \hline
          Method & $q_1$ (deg) & $q_2$ (deg) & $q_3$ (deg)& $q_4$ (deg)& $q_5$ (deg)& $q_6$ (deg) & $q_7$ (deg)\\\hline
        \multicolumn{8}{|c|}{Maximum Absolute Error $\times \paranthesis{10^{-3}}$ }\\\hline
          \makecell{Proposed\\ \policy{}} & \textbf{1} & {39} & \textbf{10} & \textbf{46} & \textbf{3.6} & \textbf{4.8} & \textbf{0.1} \\\hline
          \cite{Huang2023}  & 321 & \textbf{37} & 615 & 121 & 1850 & 450 & 1993 \\\hline
          \cite{yeCao} & 3830 & 21732 & 3347 & 17653 & 129 & 168 & 132 \\\hline
          \multicolumn{8}{|c|}{RMSE $\times \paranthesis{10^{-3}}$ }\\\hline
          \makecell{Proposed\\ \policy{}}  & \textbf{0.3} & 28& \textbf{7} & \textbf{32}& \textbf{2} & \textbf{2} & \textbf{0.0} \\\hline
          \cite{Huang2023}  & 90 & \textbf{22} & 408 & 77 & 671 & 194 & 1539 \\\hline
          \cite{yeCao} & 685 & 5134 & 483 & 2912 & 16 & 66 & 49 \\\hline
    \end{tabular}    
    \label{tab:ptcuel_iiwaPosErr}
\end{table}

\begin{table}[!t]
    \centering
    \caption{Comparison of steady-state error metrics for joint angular velocity in (degrees/second) between the proposed method \policy{}, \cite{Huang2023} and \cite{yeCao} computed after the prescribed-time $T = 4 \text{s}$ for the IIWA 14 robot.}
    \begin{tabular}{|
    >{\centering\arraybackslash}m{3em}|
    >{\centering\arraybackslash}m{2.25em}|
    >{\centering\arraybackslash}m{2.25em}|
    >{\centering\arraybackslash}m{2.25em}|
    >{\centering\arraybackslash}m{2.25em}|
    >{\centering\arraybackslash}m{2.25em}|
    >{\centering\arraybackslash}m{2.25em}|
    >{\centering\arraybackslash}m{2.25em}|}
        \hline
          Method & $\dot{q}_1 $ (deg/s) & $\dot{q}_2 $ (deg/s) & $\dot{q}_3 $ (deg/s) & $\dot{q}_4 $ (deg/s) & $\dot{q}_5 $ (deg/s) & $\dot{q}_6 $ (deg/s) & $\dot{q}_7 $ (deg/s)\\\hline
        \multicolumn{8}{|c|}{Maximum Absolute Error}\\\hline
         Proposed \policy{} & \textbf{0.2} & \textbf{0.2} & \textbf{0.40} & \textbf{0.9} & \textbf{0.4} & \textbf{2.6} & \textbf{0.05} \\\hline
           \cite{Huang2023} & 29.7 & 4.7 & 41.8 & 9.7 & 60.6 & 61.03 & 60.0 \\\hline
          \cite{yeCao} & 62.6 & 63.8 & 61.9 & 65.6 & 53.2 & 27.5 & 60.0 \\\hline
            \multicolumn{8}{|c|}{RMSE}\\\hline
          Proposed \policy{} & \textbf{0.04} & \textbf{0.07} & \textbf{0.07} & \textbf{0.26} & \textbf{0.09} & \textbf{0.81} & \textbf{0.01} \\\hline
          \cite{Huang2023} & 10.8 & 0.9 & 20.2 & 3.7 & 46.5 & 26.1 & 46.5 \\\hline
           \cite{yeCao} & 21.3 & 25.5 & 21.5 & 28.1 & 3.1 & 1.8 & 21.2 \\\hline
    \end{tabular}
    \label{tab:ptcuel_iiwaVelErr}
\end{table}

\begin{figure*}[t!]
    \centering
    \includegraphics[width=\textwidth]{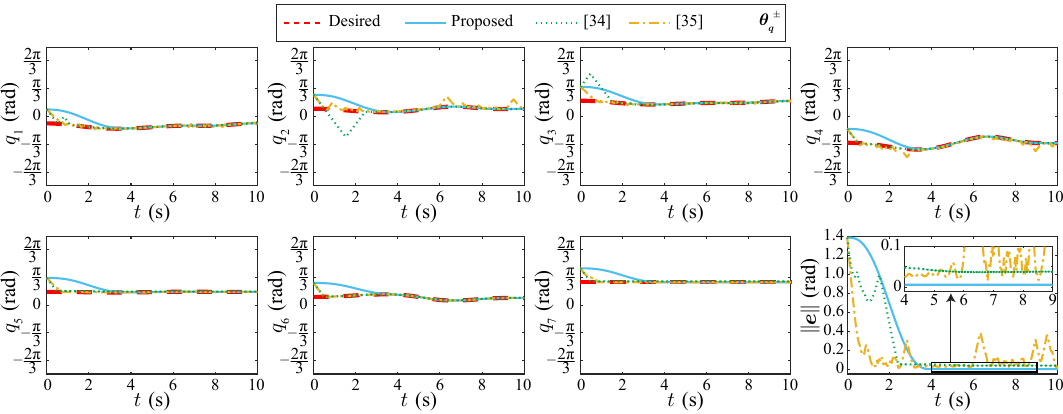}
    \caption{Simulation results for the IIWA 14 robot manipulator shows the joint angular position tracking with the prescribed-time $T = 4\text{s}$ for the proposed control scheme \policy{} against \cite{Huang2023} and \cite{yeCao}. }
    \label{fig:ptcuel_comparison_iiwa14_jointPosition}
\end{figure*}

\begin{figure*}[t!]
    \centering
    \includegraphics[width=\textwidth]{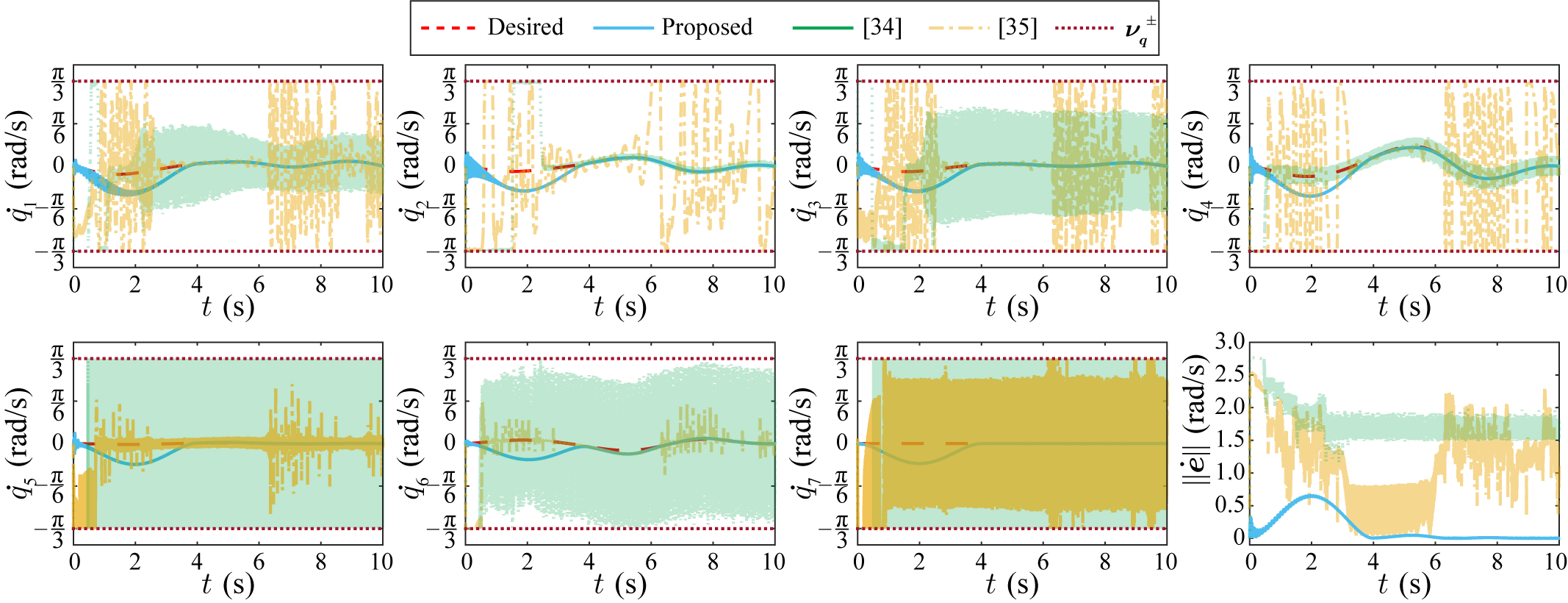}
    \caption{Simulation results for the IIWA 14 robot manipulator shows the joint angular velocity tracking with the prescribed-time $T = 4\text{s}$ for the proposed control scheme \policy{} against \cite{Huang2023} and \cite{yeCao}. }
    \label{fig:ptcuel_comparison_iiwa14_jointVelocity}
\end{figure*}

\begin{figure}[t!]
    \centering
    \includegraphics[width=0.45\textwidth]{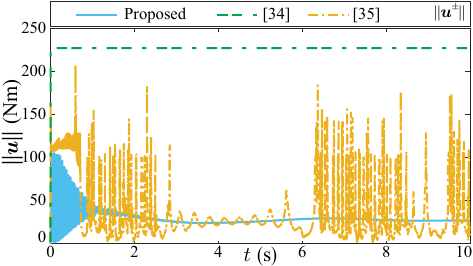}
    \caption{Simulation results for the IIWA 14 robot manipulator shows the norm of control input ($\norm{\bs{u}}$) for the proposed control scheme \policy{} against \cite{Huang2023} and \cite{yeCao}. }
    \label{fig:ptcuel_comparison_iiwa14_control_input}
\end{figure}

For the IIWA 14 robot, the error metrics in Tables \ref{tab:ptcuel_iiwaPosErr} and \ref{tab:ptcuel_iiwaVelErr} show a similar trend with both maximum absolute error and RMSE are of the order of $10^{-2}$ degrees for joint angular position error with the proposed controller \policy{}, whereas RMSE values are of the order of $10^{-1}$ degrees for the two other methods. Also, for joint angular velocity tracking, the maximum absolute error and RMSE are of the order of $10^{-1}$ degrees/second for the proposed controller, and of the order of $10$ degrees/second for the other two methods. This indicates that the proposed controller delivers significantly better performance in comparison to \cite{Huang2023} and \cite{yeCao}. This is further apparent from Fig. \ref{fig:ptcuel_comparison_iiwa14_jointPosition}, which demonstrates the significantly improved transient and steady-state performance for the proposed control scheme \policy{}. The smooth evolution of the system trajectories using the proposed scheme \policy{} arises directly from the use of the TBG functions in \eqref{eq:ptcuel_tbg_h1} and \eqref{eq:ptcuel_tbg_h2}  while inherently accounting for both state and input constraints. The contrast with alternative schemes becomes apparent from Fig. \ref{fig:ptcuel_comparison_iiwa14_jointVelocity}, where the methods in \cite{Huang2023} and \cite{yeCao} realize higher levels of chattering during velocity tracking, leading to degradation of transient and steady-state performance compared to the proposed scheme. Furthermore, the Fig. \ref{fig:ptcuel_comparison_iiwa14_control_input} shows that the control input is saturated for the method adopted from \cite{yeCao} which is impractical and the control input for the method adopted from \cite{Huang2023} has higher chatter level in comparison to the proposed policy \policy{}. Therefore, the approaches in \cite{Huang2023} and \cite{yeCao} do not scale well for higher dimensional systems, which is also apparent from the Tables \ref{tab:ptcuel_R2}, \ref{tab:ptcuel_iiwaPosErr}, and \ref{tab:ptcuel_iiwaVelErr}. A possible reason is that the controller schemes proposed in the studies \cite{Huang2023}, \cite{yeCao} do not account for the state and input constraints simultaneously, which leads to degraded performance, and these schemes implement the function approximations like RBFNN or discontinuous control policies which may be difficult to implement for higher DoF systems. 

Overall, the proposed control policy \policy{}  outperforms the other two controllers in \cite{Huang2023} and \cite{yeCao}, and achieves higher precision tracking with large initial offsets within the prescribed-time under state and input constraints in the absence of system's knowledge. 


\section{Conclusion} \label{sec:conclusion}
This study presents an approximation-free continuous adaptive barrier control strategy that achieves prescribed tracking performance for unknown EL systems subjected to state and input constraints. {In particular, the proposed scheme relies on an adaptive barrier function-based controller formulation for generating continuous and bounded control action. Moreover, stability analysis formally verifies local PTPB convergence and boundedness of the proposed time-varying gains.} Finally, in contrast with earlier studies that do not provide feasibility conditions, this {study derives sufficient conditions for minimum control authority, maximum disturbance rejection capability and a viable forward invariant set for guaranteeing local PTPB convergence.} Simulation studies are used to validate the performance of the proposed control strategy for three different robotic manipulators, thus showcasing the robustness of the controller to large variations in the parameters of the system model. A detailed quantitative comparison study with other related studies shows the superior performance of the proposed scheme. As a part of future work, the proposed scheme will be modified to directly track the trajectories in operational space when subjected to operational contact constraints for impact-aware task planning applications. In addition, the current feasibility analysis will be extended to the problem of learning feasible conditions with unknown model settings as a part of our future work. 

\appendix
\section{Proof for Lemma and Theorem} \label{FirstAppendix}
\subsection{Proof for Lemma \ref{ptcuel_Lemma}} \label{Lemma_proof}
We prove the lemma by contradiction. Choose $T<T^*$ such that control law $\bs{\mathfrak{u}}(t,\bs{x})$ guarantees PTPB convergence, i.e. $\bs{\Psi}_{\bs{\mathfrak{u}}}(t, t_0, \bs{x}(t_0))\in\mathcal{S}_\epsilon(\bs{x}_r),\  \text{for all}\ t\geq t_0+T$ , {where $\bs{x}(t_0)$ is chosen  such that distance of the point $\bs{x}(t_0) \in \mathcal{C}_2$ to the set $\mathcal{S}_\epsilon(\bs{x}_r)$ is the largest i.e.$\sup |x(t_0)|_{\mathcal{S}_\epsilon(\bs{x}_r)}$}. Then, integrating on both sides of the given control affine system from $t_0$ to $t_0+T$, we get 
\begin{flalign}
    \bs{\Psi}_{\bs{\mathfrak{u}}}(t_0{+}T, t_0, \bs{x}(t_0)) {-} \bs{x}(t_0) {=}\! \int\limits_{t_0}^{t_0+T} \!\!\!{\paranthesis{\bs{f}_a(\bs{x}) {+} \bs{g}_a(\bs{x})\bs{u}}}dt.
    \label{eq:ptcuel_integral_control_affine_a}
\end{flalign}

Now, taking the norm on both sides of \eqref{eq:ptcuel_integral_control_affine_a} and applying the bounds on $\bs{f}_a$, $\bs{g}_a$ and $\norm{\bs{u}}$, one can obtain \eqref{eq:ptcuel_lemma_1_contradiction}. Then, \eqref{eq:ptcuel_lemma_1_contradiction} implies that the solution remains outside the set $\mathcal{S}_{\epsilon}(\bs{x}_r)$ for prescribed time $T$ chosen lesser than $T^*$, which contradicts the aforementioned statement. {Thus, prescribed time $T$ has to be chosen larger than $T^*$ for arbitrary initial states started in the set $\mathcal{C}_2$ for a given feedback control law $\bs{u}$ to guarantee local PTPB convergence as in the \emph{Definition} \ref{def:ptpb_def}  under input constraints}.
\begin{flalign}
    \norm{\bs{\Psi}_{\bs{\mathfrak{u}}}&(t_0{+}T, t_0, \bs{x}(t_0)) {-} \bs{x}(t_0)}\nonumber\\ &{\leq} \!\int\limits_{t_0}^{t_0{+}T} \!\!\norm{\paranthesis{\bs{f}_a(\bs{x}) {+} \bs{g}_a(\bs{x})\bs{u}}}dt 
     \leq \int\limits_{t_0}^{t_0{+}T} \!\!(\ol{f} {+} \ol{g}u^*)dt  \nonumber\\
    & {=} (\ol{f}{+} \ol{g}u^*)T< (\ol{f} + \ol{g}u^*)T^* = |\bs{x}(t_0)|_{\mathcal{S}_\epsilon(\bs{x}_r)}.
    \label{eq:ptcuel_lemma_1_contradiction}
\end{flalign}
\subsection{Proof for Theorem \ref{ptcuel_feasibility_theorem}} \label{ptcuel_feasibility_theorem_proof}
Without loss of generality, take $\alpha_i(a_i) = a_i,\ \forall i \in \N_N$ in \eqref{eq:ptcuel_higher_order}, which implies $\bs{\alpha}{\paranthesis{\ol{\bs{\zeta}}}} = \ol{\bs{\zeta}}$ and $\bs{\alpha}{\paranthesis{\udl{\bs{\zeta}}}} = \udl{\bs{\zeta}}$. Since $\q^*$ is a constant, it results in $\dqd = \bs{0}_n,\ \dqdd = \bs{0}_n$ . Then \eqref{eq:ptcuel_higher_order} becomes,
\begin{equation}
    \ol{\bs{\beta}}(\bs{x}) = -\epsd + \sigma\ol{\bs{\zeta}}(\q),\, \udl{\bs{\beta}}(\bs{x}) = \epsd + \sigma\udl{\bs{\zeta}}(\bs{\q}).
    \label{eq:ptcuel_modified_higher_order}
\end{equation}

Now by invoking the forward invariance properties of the set $\C_4$, using Nagumo's theorem \cite{CBF_SC, blanchini1999set}, we have
\begin{equation}
    \begin{aligned}
    \dot{\ol{\bs{\beta}}}(\bs{x}) &= -\qdd + \ddot{\err}^d + \sigma\dot{\ol{\bs{\zeta}}}(\q) \succeq \bs{0}_n ,\\ \dot{\udl{\bs{\beta}}}(\bs{x}) &= \qdd  - \ddot{\err}^d + \sigma\dot{\udl{\bs{\zeta}}}(\q) \succeq \bs{0}_n .
    \end{aligned}
    \label{eq:ptcuel_higher_order_derivative}
\end{equation}
Using \eqref{eq:ptcuel_first_order}, the inequality \eqref{eq:ptcuel_higher_order_derivative} can be written as:
\begin{flalign}
    -\qdd + \ddot{\err}^d - \sigma\qd \succeq &\bs{0}_n,  \ 
    \qdd - \ddot{\err}^d + \sigma\qd \succeq \bs{0}_n, \nonumber 
\end{flalign}
From the above inequality, one can conclude the following:
\begin{flalign}
    \qdd &= \ddot{\err}^d - \sigma\qd .
     \label{eq:ptcuel_min_ctrl_auth_bare_eqn}
\end{flalign}
Now, using \eqref{eq:ptcuel_eom}, \eqref{eq:ptcuel_min_ctrl_auth_bare_eqn} and neglecting external disturbances, one can arrive at the following:
\begin{equation}
    \bs{M}^{-1}\bs{u} = \bs{M}^{-1}\paranthesis{\bs{C}\qd+\bs{G}+\bs{F} + \sigma\bs{M}\qd - \bs{M}\ddot{\err}^d} .
    \label{eq:ptcuel_min_ctrl_auth_dyn}
\end{equation}
Since $\norm{\bs{M}^{-1}\bs{u}}\leq\norm{\bs{M}^{-1}}\norm{\bs{u}}\leq\ol{m}\norm{\bs{u}}$, then using the \textit{Property} \ref{prop:mass}, followed by taking the norm on both sides of \eqref{eq:ptcuel_min_ctrl_auth_dyn}, one can obtain the following inequality:
\begin{flalign}
    \ol{m} \norm{\bs{u}} {\geq} \norm{{\bs{M}^{-1}}\paranthesis{\bs{C}\qd{+}\bs{G}{+}\bs{F} {-} \sigma\bs{M}\qd {+} \bs{M}\ddot{\err}^d}}.\label{eq:ptcuel_min_control_ineq_lower}
\end{flalign}
In order to obtain the lower bound on the minimum control authority, we consider the uniform ultimate bound on the right-hand side of the inequality \eqref{eq:ptcuel_min_control_ineq_lower} by using the \textit{Properties} \ref{prop:mass}, \ref{prop:matrix_bounds} and state constraints as in \eqref{eq:ptcuel_min_control_authority_inequality}.
\begin{flalign}
   &\norm{{\bs{M}^{-1}}\paranthesis{\bs{C}\qd{+}\bs{G}{+}\bs{F} {-} \sigma\bs{M}\qd {+} \bs{M}\ddot{\err}^d}} \leq \nonumber\\&\ol{m}\paranthesis{\ol{C}(\nu_q^+)^2+\ol{F}\nu_q^++\sigma \ol{M}\nu_q^+ + \ol{G} + \ol{M}\ol{e}_{dd}}. \label{eq:ptcuel_min_control_authority_inequality}
\end{flalign}
Then, using \eqref{eq:ptcuel_min_control_ineq_lower} and \eqref{eq:ptcuel_min_control_authority_inequality}, we get
\begin{flalign}
    \norm{\bs{u}} \geq  \paranthesis{\ol{C}(\nu_q^+)^2{+}\ol{F}\nu_q^+{+}\sigma \ol{M}\nu_q^+ {+} \ol{G} {+} \ol{M}\ol{e}_{dd}}.
    \label{eq:ptcuel_min_viable_base}
\end{flalign}
Then, by solving \eqref{eq:ptcuel_min_viable_base} for $\ol{e}_{dd}$ and using \eqref{eq:ptcuel_tbg_bounds} and the expression for $u^*$, one can obtain the set ${\Set}$ in \eqref{eq:ptcuel_analytical_viable_set}. Additionally, if $\sigma\geq\ol{\sigma}$ then, one can deduce ${\Set} = \varnothing$. Therefore, $\sigma$ is chosen within the range $(\udl{\sigma}, \ol{\sigma})$ to render the viable set ${\Set}$. Also, note that the choice of $T>2\Max\paranthesis{\norm{\bs{x}^+}, \norm{\bs{x}^-}}/(\ol{f}+\ol{g}u^*)>T^*$ ensures that inequality \eqref{eq:ptcuel_PT_lower_bound} is satisfied so that for any initial condition starting in ${\Set}$, PTC is guaranteed. 

Lastly, the minimum control authority $u_\Min$ is then directly obtained by replacing the variable $\ol{e}_{dd}$ with its maximum value in \eqref{eq:ptcuel_edd_max}. Correspondingly, the maximum disturbance $\ol{d}$ that the controller can handle may be directly derived as $\ol{d}=u^*-u_\Min$ resulting in \eqref{eq:ptcuel_max_disturbances}. \hfill $\blacksquare$


\begin{thebibliography}{10}
\providecommand{\url}[1]{#1}
\csname url@samestyle\endcsname
\providecommand{\newblock}{\relax}
\providecommand{\bibinfo}[2]{#2}
\providecommand{\BIBentrySTDinterwordspacing}{\spaceskip=0pt\relax}
\providecommand{\BIBentryALTinterwordstretchfactor}{4}
\providecommand{\BIBentryALTinterwordspacing}{\spaceskip=\fontdimen2\font plus
\BIBentryALTinterwordstretchfactor\fontdimen3\font minus \fontdimen4\font\relax}
\providecommand{\BIBforeignlanguage}[2]{{%
\expandafter\ifx\csname l@#1\endcsname\relax
\typeout{** WARNING: IEEEtran.bst: No hyphenation pattern has been}%
\typeout{** loaded for the language `#1'. Using the pattern for}%
\typeout{** the default language instead.}%
\else
\language=\csname l@#1\endcsname
\fi
#2}}
\providecommand{\BIBdecl}{\relax}
\BIBdecl

\bibitem{ogata2010modern}
K.~Ogata, \emph{Modern Control Engineering}, 4th~ed.\hskip 1em plus 0.5em minus 0.4em\relax USA: Prentice Hall PTR, 2001.

\bibitem{Shen_FT_OBSER}
Y.~Shen and X.~Xia, ``Semi-global finite-time observers for nonlinear systems,'' \emph{Automatica}, vol.~44, no.~12, pp. 3152--3156, 2008.

\bibitem{basin_FT}
M.~Basin, ``Finite- and fixed-time convergent algorithms: Design and convergence time estimation,'' \emph{Annual Reviews in Control}, vol.~48, pp. 209--221, 2019.

\bibitem{polykao_FT_EXPAP}
F.~Lopez-Ramirez, D.~Efimov, A.~Polyakov, and W.~Perruquetti, ``Finite-time and fixed-time input-to-state stability: Explicit and implicit approaches,'' \emph{Systems \& Control Letters}, vol. 144, p. 104775, 2020.

\bibitem{polykov_FT_NON}
A.~Polyakov, ``Nonlinear feedback design for fixed-time stabilization of linear control systems,'' \emph{IEEE Transactions on Automatic Control}, vol.~57, no.~8, pp. 2106--2110, 2012.

\bibitem{PTS_IMA}
J.~D. Sánchez-Torres, D.~Gómez-Gutiérrez, E.~López, and A.~G. Loukianov, ``{A class of predefined-time stable dynamical systems},'' \emph{IMA Journal of Mathematical Control and Information}, vol.~35, no. Supplement\_1, pp. i1--i29, 02 2017.

\bibitem{PTC_note_ifac}
E.~Jiménez-Rodríguez, A.~J. Muñoz-Vázquez, J.~D. Sánchez-Torres, and A.~G. Loukianov, ``A note on predefined-time stability,'' \emph{IFAC-PapersOnLine}, vol.~51, no.~13, pp. 520--525, 2018.

\bibitem{obuz:2025}
S.~Obuz, E.~Selim, E.~Tatlicioglu, and E.~Zergeroglu, ``Robust prescribed time
  control of euler–lagrange systems,'' \emph{IEEE Transactions on Industrial
  Electronics}, vol.~72, no.~2, pp. 1694--1701, 2025.

\bibitem{song:2023}
Y.~Song, H.~Ye, and F.~L. Lewis, ``Prescribed-time control and its latest developments,'' \emph{IEEE Transactions on Systems, Man, and Cybernetics: Systems}, vol.~53, no.~7, pp. 4102--4116, 2023.

\bibitem{CBF_EL_IP}
W.~S. Cortez and D.~V. Dimarogonas, ``Correct-by-design control barrier functions for euler-lagrange systems with input constraints,'' in \emph{2020 American Control Conference (ACC)}, 2020, pp. 950--955.

\bibitem{yang_adative}
C.~Yang, Y.~Jiang, W.~He, J.~Na, Z.~Li, and B.~Xu, ``Adaptive parameter estimation and control design for robot manipulators with finite-time convergence,'' \emph{IEEE Transactions on Industrial Electronics}, vol.~65, no.~10, pp. 8112--8123, 2018.

\bibitem{sun_fxd_ppf}
Y.~Sun, J.~Kuang, Y.~Gao, W.~Chen, J.~Wang, J.~Liu, and L.~Wu, ``Fixed-time prescribed performance tracking control for manipulators against input saturation,'' \emph{Nonlinear Dynamics}, vol. 111, no.~15, pp. 14\,077--14\,095, Aug. 2023.

\bibitem{bertingo_7dof}
A.~Bertino, P.~Naseradinmousavi, and M.~Krstic, ``Experimental and analytical prescribed-time trajectory tracking control of a 7-dof robot manipulator,'' in \emph{2022 American Control Conference (ACC)}, 2022, pp. 1941--1946.

\bibitem{PTC_TBG}
G.~Arechavaleta, J.~Obreg{\'o}n, H.~M. Becerra, and A.~Morales-D{\'\i}az, ``Predefined-time convergence in task-based inverse dynamics using time base generators,'' \emph{IFAC-PapersOnLine}, vol.~51, no.~13, pp. 443--449, 2018.

\bibitem{song:2017}
Y.~Song, Y.~Wang, J.~Holloway, and M.~Krstic, ``Time-varying feedback for regulation of normal-form nonlinear systems in prescribed finite time,'' \emph{Automatica}, vol.~83, pp. 243--251, 2017.

\bibitem{holloway:linear_observers:2019}
J.~Holloway and M.~Krstic, ``Prescribed-time observers for linear systems in observer canonical form,'' \emph{IEEE Transactions on Automatic Control}, vol.~64, no.~9, pp. 3905--3912, 2019.

\bibitem{hua:aptc:2022}
C.~Hua, P.~Ning, and K.~Li, ``Adaptive prescribed-time control for a class of uncertain nonlinear systems,'' \emph{IEEE Transactions on Automatic Control}, vol.~67, no.~11, pp. 6159--6166, 2022.

\bibitem{hua:time_dealy:2023}
C.~Hua, H.~Li, K.~Li, and P.~Ning, ``Adaptive prescribed-time control of time-delay nonlinear systems via a double time-varying gain approach,'' \emph{IEEE Transactions on Cybernetics}, vol.~53, no.~8, pp. 5290--5298, 2023.

\bibitem{krishnaMurthy:dynamicHighGain:2020}
P.~Krishnamurthy, F.~Khorrami, and M.~Krstic, ``A dynamic high-gain design for prescribed-time regulation of nonlinear systems,'' \emph{Automatica}, vol. 115, p. 108860, 2020.

\bibitem{PTC_perturbed_EL}
A.~Shakouri and N.~Assadian, ``Prescribed-time control for perturbed euler-lagrange systems with obstacle avoidance,'' \emph{IEEE Transactions on Automatic Control}, vol.~67, no.~7, pp. 3754--3761, 2022.

\bibitem{kgarg_clf}
K.~Garg, E.~Arabi, and D.~Panagou, ``Prescribed-time convergence with input constraints: A control lyapunov function based approach,'' in \emph{2020 American Control Conference (ACC)}, 2020, pp. 962--967.

\bibitem{Huang2021-ep}
K.-L. Huang, M.-F. Ge, C.-D. Liang, J.-W. Dong, and X.-W. Zhao, ``Hierarchical predefined-time control for time-varying formation tracking of multiple heterogeneous {Euler--Lagrange} agents,'' \emph{Nonlinear Dynamics}, vol. 105, no.~4, pp. 3255--3270, Sep. 2021.

\bibitem{denis:discretization:2024}
D.~Efimov and Y.~Orlov, ``Discretization of prescribed-time observers in the presence of noises and perturbations,'' \emph{Systems \& Control Letters}, vol. 188, p. 105820, 2024.

\bibitem{yorlov:timeSpace:2022}
Y.~Orlov, ``Time space deformation approach to prescribed-time stabilization: Synergy of time-varying and non-lipschitz feedback designs,'' \emph{Automatica}, vol. 144, p. 110485, 2022.

\bibitem{gutierrez_exact_diff:2023}
D.~Gómez-Gutiérrez, R.~Aldana-López, R.~Seeber, M.~T. Angulo, and L.~Fridman, ``An arbitrary-order exact differentiator with predefined convergence time bound for signals with exponential growth bound,'' \emph{Automatica}, vol. 153, p. 110995, 2023.

\bibitem{yorlov:2022:robust_differentiators}
Y.~Orlov, R.~I. Verdés~Kairuz, and L.~T. Aguilar, ``Prescribed-time robust differentiator design using finite varying gains,'' \emph{IEEE Control Systems Letters}, vol.~6, pp. 620--625, 2022.

\bibitem{ramon:robust_obser_2022}
R.~I. {Verdés Kairuz}, Y.~Orlov, and L.~T. Aguilar, ``Robust observer design with prescribed settling-time bound and finite varying gains,'' \emph{European Journal of Control}, vol.~68, p. 100667, 2022, 2022 European Control Conference Special Issue.

\bibitem{silva:robust_sensor_noise:2024}
J.~F. Silva and D.~A. Santos, ``On the robustness of a modified super-twisting algorithm with prescribed-time convergence,'' \emph{IEEE Access}, vol.~12, pp. 58\,106--58\,113, 2024.

\bibitem{lopez_diff:2023}
R.~Aldana-López, R.~Seeber, D.~Gómez-Gutiérrez, M.~T. Angulo, and M.~Defoort, ``A redesign methodology generating predefined-time differentiators with bounded time-varying gains,'' \emph{International Journal of Robust and Nonlinear Control}, vol.~33, no.~15, pp. 9050--9065, 2023.

\bibitem{zhao_PPC}
K.~Zhao, Y.~Song, T.~Ma, and L.~He, ``Prescribed performance control of uncertain euler–lagrange systems subject to full-state constraints,'' \emph{IEEE Transactions on Neural Networks and Learning Systems}, vol.~29, no.~8, pp. 3478--3489, 2018.

\bibitem{sun_Neural}
Y.~Sun, Y.~Gao, Y.~Zhao, Z.~Liu, J.~Wang, J.~Kuang, F.~Yan, and J.~Liu, ``Neural network-based tracking control of uncertain robotic systems: Predefined-time nonsingular terminal sliding-mode approach,'' \emph{IEEE Transactions on Industrial Electronics}, vol.~69, no.~10, pp. 10\,510--10\,520, 2022.

\bibitem{robustPPC_EL_PTC}
Z.~Yin, J.~Luo, and C.~Wei, ``Robust prescribed performance control for euler–lagrange systems with practically finite-time stability,'' \emph{European Journal of Control}, vol.~52, pp. 1--10, 2020.

\bibitem{PTC_EL_2020_Jian}
J.~Li, J.~Li, Z.~Wu, and Y.~Liu, ``Practical tracking control with prescribed transient performance for euler-lagrange equation,'' \emph{Journal of the Franklin Institute}, vol. 357, no.~10, pp. 5809--5830, 2020.

\bibitem{PPC_BLF_EL}
J.~Zhang, J.~Yang, Z.~Zhang, and Y.~Wu, ``Prescribed performance control of euler–lagrange systems with input saturation and output constraints,'' \emph{International Journal of Adaptive Control and Signal Processing}, vol.~36, no.~12, pp. 3184--3204, 2022.

\bibitem{Huang2023}
X.-W. Huang, Z.-Y. Dong, P.~Yang, and L.-H. Zhang, ``Model-free adaptive trajectory tracking control of robotic manipulators with practical prescribed-time performance,'' \emph{Nonlinear Dynamics}, Sep 2023.

\bibitem{yeCao}
Y.~Cao, J.~Cao, and Y.~Song, ``Practical prescribed time control of euler–lagrange systems with partial/full state constraints: A settling time regulator-based approach,'' \emph{IEEE Transactions on Cybernetics}, vol.~52, no.~12, pp. 13\,096--13\,105, 2022.

\bibitem{pptc_song}
Z.~Song and K.~Sun, ``Prescribed performance tracking control for a class of nonlinear system considering input and state constraints,'' \emph{ISA transactions}, vol. 119, pp. 81--92, 2022.

\bibitem{lopez_noise:2023}
R.~Aldana-López, R.~Seeber, H.~Haimovich, and D.~Gómez-Gutiérrez, ``On inherent limitations in robustness and performance for a class of prescribed-time algorithms,'' \emph{Automatica}, vol. 158, p. 111284, 2023.

\bibitem{kgarg_qp}
K.~Garg and D.~Panagou, ``Control-lyapunov and control-barrier functions based quadratic program for spatio-temporal specifications,'' in \emph{2019 IEEE 58th Conference on Decision and Control (CDC)}, 2019, pp. 1422--1429.

\bibitem{ehsan_TimeTF}
E.~Arabi, T.~Yucelen, and J.~R. Singler, ``Robustness of finite-time distributed control algorithm with time transformation,'' in \emph{2019 American Control Conference (ACC)}, 2019, pp. 108--113.

\bibitem{khalil}
H.~K. Khalil, \emph{Nonlinear Systems}.\hskip 1em plus 0.5em minus 0.4em\relax
  Englewood Cliffs, NJ, USA: Prentice-Hall, 2002.
  
\bibitem{TBG_Higher_order}
H.~M. Becerra, C.~R. Vázquez, G.~Arechavaleta, and J.~Delfin, ``Predefined-time convergence control for high-order integrator systems using time base generators,'' \emph{IEEE Transactions on Control Systems Technology}, vol.~26, no.~5, pp. 1866--1873, 2018.

\bibitem{haifeg_PTC_ADA}
H.~Ma, W.~Liu, Z.~Xiong, Y.~Li, Z.~Liu, and Y.~Sun, ``Predefined-time barrier function adaptive sliding-mode control and its application to piezoelectric actuators,'' \emph{IEEE Transactions on Industrial Informatics}, vol.~18, no.~12, pp. 8682--8691, 2022.

\bibitem{zhang_backstepping}
L.~Zhang, J.~Liu, and N.~Cui, ``Backstepping control for a two-link manipulator with appointed-time convergence,'' \emph{ISA Transactions}, vol. 128, pp. 208--219, 2022.

\bibitem{bruno_book}
B.~Siciliano, L.~Sciavicco, L.~Villani, and G.~Oriolo, \emph{Robotics: Modelling, Planning and Control}.\hskip 1em plus 0.5em minus 0.4em\relax Springer Publishing Company, Incorporated, 2010.

\bibitem{EL_props}
R.~Ortega, A.~Lor{\'i}a, P.~J. Nicklasson, and H.~Sira-Ram{\'i}rez, \emph{Euler-Lagrange systems}.\hskip 1em plus 0.5em minus 0.4em\relax London: Springer London, 1998, pp. 15--37.

\bibitem{van_fault_tolerant}
M.~Van, S.~S. Ge, and H.~Ren, ``Finite time fault tolerant control for robot manipulators using time delay estimation and continuous nonsingular fast terminal sliding mode control,'' \emph{IEEE Transactions on Cybernetics}, vol.~47, no.~7, pp. 1681--1693, 2017.

\bibitem{sun_USDE}
C.~Sun, S.~Wang, and H.~Yu, ``Finite-time sliding mode control based on unknown system dynamics estimator for nonlinear robotic systems,'' \emph{IEEE Transactions on Circuits and Systems II: Express Briefs}, vol.~70, no.~7, pp. 2535--2539, 2023.

\bibitem{xiao_UD}
B.~Xiao, S.~Yin, and O.~Kaynak, ``Tracking control of robotic manipulators with uncertain kinematics and dynamics,'' \emph{IEEE Transactions on Industrial Electronics}, vol.~63, no.~10, pp. 6439--6449, 2016.

\bibitem{wu_adaptive_rbfnn}
Y.~Wu, H.~Fang, T.~Xu, and F.~Wan, ``Adaptive neural fixed-time sliding mode control of uncertain robotic manipulators with input saturation and prescribed constraints,'' \emph{Neural Processing Letters}, vol.~54, no.~5, pp. 3829--3849, Oct. 2022.

\bibitem{wu_fxd_neural}
Y.~Wu, W.~Niu, L.~Kong, X.~Yu, and W.~He, ``Fixed-time neural network control of a robotic manipulator with input deadzone,'' \emph{ISA Transactions}, vol. 135, pp. 449--461, 2023.

\bibitem{yu_FxdTC}
Z.~Yu, K.~Pang, and C.~Hua, ``Global composite learning fixed-time control of robotic systems with asymmetric tracking error constraints,'' \emph{IEEE Transactions on Industrial Electronics}, vol.~70, no.~7, pp. 7082--7091, 2023.

\bibitem{sun_PPC}
L.~Sun, H.~Cao, and Y.~Song, ``Prescribed performance control of constrained euler–language systems chasing unknown targets,'' \emph{IEEE Transactions on Cybernetics}, vol.~53, no.~8, pp. 4829--4840, 2023.

\bibitem{chen:2023:eventDriven}
Z.~Chen, H.~Zhang, J.~Liu, Q.~Wang, and J.~Wang, ``Adaptive prescribed settling time periodic event-triggered control for uncertain robotic manipulators with state constraints,'' \emph{Neural Networks}, vol. 166, pp. 1--10, 2023.

\bibitem{hefu:timeVarying:2023}
H.~Ye and Y.~Song, ``Prescribed-time control for time-varying nonlinear systems: A temporal scaling based robust adaptive approach,'' \emph{Systems \& Control Letters}, vol. 181, p. 105602, 2023.

\bibitem{amir:2022:timevarying}
A.~Shakouri, ``On the prescribed-time attractivity and frozen-time eigenvalues of linear time-varying systems,'' \emph{Automatica}, vol. 140, p. 110173, 2022.

\bibitem{wuquan:vanishing:2024}
W.~Li and M.~Krstic, ``Prescribed-time control of nonlinear systems with linearly vanishing multiplicative measurement noise,'' \emph{IEEE Transactions on Automatic Control}, vol.~69, no.~6, pp. 3647--3661, 2024.

\bibitem{quad_EL}
S.~Martini, S.~Sönmez, A.~Rizzo, M.~Stefanovic, M.~J. Rutherford, and K.~P. Valavanis, ``Euler-lagrange modeling and control of quadrotor uav with aerodynamic compensation,'' in \emph{2022 International Conference on Unmanned Aircraft Systems (ICUAS)}, 2022, pp. 369--377.

\bibitem{kgarg_FxTsInput:2022}
K.~Garg, E.~Arabi, and D.~Panagou, ``Fixed-time control under spatiotemporal and input constraints: A quadratic programming based approach,'' \emph{Automatica}, vol. 141, p. 110314, 2022.

\bibitem{Robot_Control:1990}
S.~Nicosia and P.~Tomei, ``Robot control by using only joint position measurements,'' \emph{IEEE Transactions on Automatic Control}, vol.~35, no.~9, pp. 1058--1061, 1990.

\bibitem{ZHANG2008475}
Y.~Zhang, X.~Lv, Z.~Li, Z.~Yang, and K.~Chen, ``Repetitive motion planning of pa10 robot arm subject to joint physical limits and using lvi-based primal–dual neural network,'' \emph{Mechatronics}, vol.~18, no.~9, pp. 475--485, 2008.

\bibitem{CBF_SC}
A.~D. Ames, X.~Xu, J.~W. Grizzle, and P.~Tabuada, ``Control barrier function based quadratic programs for safety critical systems,'' \emph{IEEE Transactions on Automatic Control}, vol.~62, no.~8, pp. 3861--3876, 2017.

\bibitem{setTheoretic}
S.~M. Franco~Blanchini, \emph{Set Theoretic in Control}.\hskip 1em plus 0.5em minus 0.4em\relax Birkhäuser Boston, MA, 2007.

\bibitem{barrierFunc}
K.~Shao, J.~Zheng, R.~Tang, X.~Li, Z.~Man, and B.~Liang, ``Barrier function based adaptive sliding mode control for uncertain systems with input saturation,'' \emph{IEEE/ASME Transactions on Mechatronics}, vol.~27, no.~6, pp. 4258--4268, 2022.

\bibitem{rudin}
W.~Rudin, \emph{Principles of Mathematical Analysis}, 3rd~ed.\hskip 1em plus 0.5em minus 0.4em\relax McGraw-Hill, 1976.

\bibitem{funnel_mobile_el}
F.~Mehdifar, C.~P. Bechlioulis, and D.~V. Dimarogonas, ``Funnel control under hard and soft output constraints,'' in \emph{2022 IEEE 61st Conference on Decision and Control (CDC)}, 2022, pp. 4473--4478.

\bibitem{Obeid:2018}
H.~Obeid, L.~M. Fridman, S.~Laghrouche, and M.~Harmouche, ``Barrier function-based adaptive sliding mode control,'' \emph{Automatica}, vol.~93, pp. 540--544, 2018.

\bibitem{Obeid:2020}
H.~Obeid, S.~Laghrouche, L.~Fridman, Y.~Chitour, and M.~Harmouche, ``Barrier function-based adaptive super-twisting controller,'' \emph{IEEE Transactions on Automatic Control}, vol.~65, no.~11, pp. 4928--4933, 2020.

\bibitem{adaptivePI:2017}
Y.~Song, Y.~Wang, and C.~Wen, ``Adaptive fault-tolerant pi tracking control with guaranteed transient and steady-state performance,'' \emph{IEEE Transactions on Automatic Control}, vol.~62, no.~1, pp. 481--487, 2017.

\bibitem{rajpal_ZNN}
R.~Singh and J.~Keshavan, ``A provably constrained neural control architecture with prescribed performance for fault-tolerant redundant manipulators,'' \emph{IEEE Access}, vol.~10, pp. 97\,719--97\,732, 2022.

\bibitem{blanchini1999set}
F.~Blanchini, ``Set invariance in control,'' \emph{Automatica}, vol.~35, no.~11, pp. 1747--1767, 1999.

\end{thebibliography}
\end{document}